\documentclass[prl,twocolumn,superscriptaddress,amsmath,amssymb,floatfix,noshowpacs]{revtex4}
\usepackage{graphicx,hyperref}
\usepackage{color}
\usepackage{soul}
\usepackage{balance}
\usepackage{amsmath}
\usepackage{tabularx}
\graphicspath{{fig/}}

\newcommand*{\ket}[1]{\vert #1 \rangle}

\newcommand{\hide}[1]{}

\begin{document}
\title{\textbf{Multi-channel waveguide QED with atomic arrays in free space}}
\author{Yakov Solomons}
\affiliation{Department of Chemical \& Biological Physics, Weizmann Institute of Science, Rehovot 7610001, Israel}
\author{Ephraim Shahmoon}
\affiliation{Department of Chemical \& Biological Physics, Weizmann Institute of Science, Rehovot 7610001, Israel}
\date{\today}

\begin{abstract}
We study light scattering off a two-dimensional (2D) array of atoms driven to Rydberg levels. We show that the problem can be mapped to a generalized model of waveguide QED, consisting of multiple 1D photonic channels (transverse modes), each of which directionally coupled to a corresponding Rydberg surface mode of the array.
In the Rydberg blockade regime, collective excitations of different surface modes block each other, leading to multi-channel correlated photonic states. Using an analytical approach, we characterize inter-channel quantum correlations, and elucidate the role of collective two-photon resonances of the array. Our results open new possibilities for multimode many-body physics and quantum information with photons in a free-space platform.
\end{abstract}
 \maketitle

The exploration of quantum optical science and technology relies on physical platforms which allow for efficient interfaces between light and matter. In particular, these typically require: (i) strong coupling between matter, e.g. atoms, and a well-defined, directional photon mode; and (ii) nonlinearity of the atomic media which can generate non-trivial quantum correlations. A prominent paradigm for such quantum optical platforms is that of waveguide QED, where dielectric structures tightly confine and guide light, enabling its strong and directional coupling to atoms trapped nearby \cite{CHAr2,KIMc,RAU1,RAU2,KIM2,OROZ,LOD,TOM}. Nonlinear photon-atom scattering in such systems can then entangle the photons, leading to applications in quantum information and technology \cite{CHAr1,DK,BARg,CHA,KIMd,DAY1,LUK4,REM1,CIR1,ALJ3,REM2}. Crucially for atomic systems, this approach relies on the demanding task of trapping atoms very close to the waveguiding dielectric structures. Alternatively, strong light-matter interaction can be achieved for light propagating inside large atomic ensembles trapped in free space \cite{POLr,FIM}, though applications such as quantum gates are typically limited by the effective interaction length inside the atomic medium \cite{FIRr,GOR1,FRI,FIR}.

\begin{figure}[t]
  \begin{center}
    \includegraphics[width=\columnwidth]{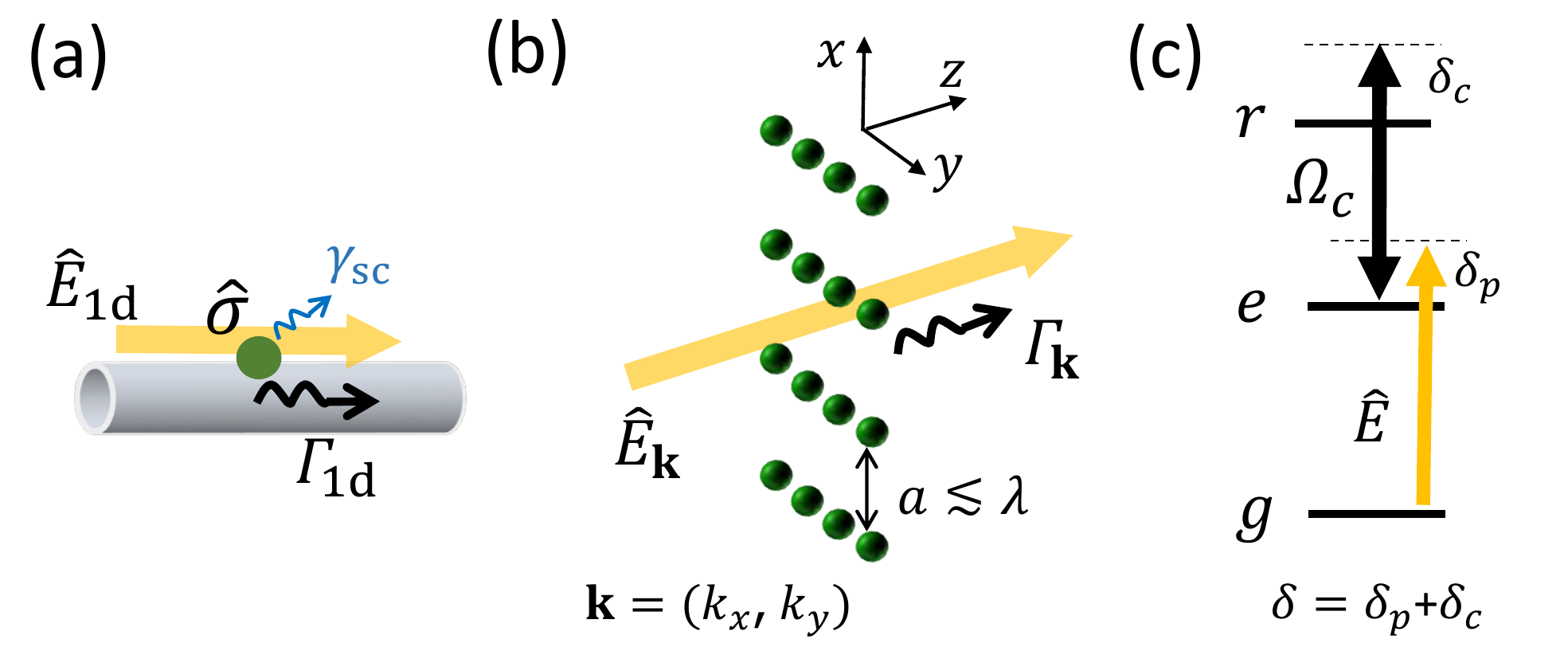}
    \caption{\small{
(a) Standard waveguide QED setup: a two-level atom ($\hat{\sigma}$) is coupled to a 1D mode $\hat{E}_{\mathrm{1d}}$ at rate $\Gamma_{\mathrm{1d}}$, and to non-guided modes at rate $\gamma_{\mathrm{sc}}$. (b) 2D atom array setup: atoms span the $xy$ plane with near-wavelength lattice spacing $a$. Conservation of lattice momentum $\mathbf{k}$ leads to directional coupling to transverse photon modes $\hat{E}_{\mathbf{k}}$ at rate $\Gamma_{\mathbf{k}}$. (c) Level structure of an array atom.
    }} \label{fig1}
  \end{center}
\end{figure}

Here instead, we discuss a free-space platform which is based on light scattering off a single 2D atomic array, in analogy to the scattering off an atom in waveguide QED (Fig. 1). Specifically, we consider free-space, propagating photon modes and their interaction with a 2D array of laser-trapped atoms with near-wavelength lattice-spacing, realizable e.g. in an optical lattice \cite{OL1,OL2}. Light-matter coupling in this system is extremely strong and directional, as can be characterized by the high reflectivity of the array. The latter has been recently predicted \cite{ADM,coop} and observed \cite{RUI}, followed by various proposals for applications in quantum science and technology \cite{RUO2,MNZ,ZOL1,ZOL2,HEN,AAMO,om1,om2,RUO3,ANA,ANAm,TAY,CIRw,RITc,cQED,JAN1,ROB2}. When the array atoms are additionally excited to Rydberg states, the resulting strong interactions between atoms can be exploited for quantum information processing \cite{LUKr,RIV,CIR,CHA4,POH4}.

We show that the combination of the strong directional light-matter coupling together with the nonlinearity associated with Rydberg excitations gives rise to generalized waveguide QED physics described by multiple 1D photonic ``channels". Each 1D channel corresponds to a different propagating transverse mode, e.g. a beam at a certain incident angle, which is directionally coupled to a spatially matched Rydberg surface excitation of the 2D atom array. Interactions between these collective excitations render them highly nonlinear, leading to directional, multi-channel nonlinear scattering. We find analytically that quantum correlations between scattered photons of different channels are strongest at a collective two-photon resonance of the Rydberg surface excitations. Moreover, the figure of merit that characterizes the strength of correlations is the array reflectivity.
These results open the prospect for studying a novel, multimode generalization of waveguide QED physics, which does not exist in typical waveguide systems.

\emph{System.---}
We consider a 2D array of three-level atoms spanning the $xy$ plane (Figs. 1b,c). The ground ($|g\rangle$) to excited ($|e\rangle$) atomic transition is coupled to a weak quantum field $\hat{E}(\mathbf{r},t)=\sum_{\mathbf{k}}\hat{E}_{\mathbf{k}}(z,t)e^{i\mathbf{k}\cdot\mathbf{r}}$, spanned by its components of different transverse (in-plane) wavevectors $\mathbf{k}=(k_x,k_y)$, and whose average coherent input oscillates at frequency $\omega_p$ detuned by $\delta_p$ from the $|g\rangle\leftrightarrow |e\rangle$ transition. In addition, the level $|e\rangle$ of each atom is coupled to a highly excited, long-lived Rydberg level $|r\rangle$ via a coupling field with Rabi frequency $\Omega_c$ and detuning $\delta_c$ from the $|e\rangle\leftrightarrow |r\rangle$ transition.
For an array of two-level atoms ($\Omega_c=0$) with near-wavelength lattice spacing, $a\lesssim \lambda=2\pi c/\omega_p$, photon-mediated dipole-dipole interactions between the different atoms become significant \cite{BRW1,BRW2}: they lead to the formation of collective dipole modes, which by translational invariance are characterized by lattice momentum $\mathbf{k}$. The resonant frequencies of these collective modes are cooperatively shifted by $\Delta_{\mathbf{k}}$ from that of an isolated atom, and are only coupled to propagating photons with the same in-plane wavevector $\mathbf{k}$, at a modified cooperative emission rate $\Gamma_{\mathbf{k}}$ \cite{coop} (Fig. 1b).

\emph{Rydberg surface modes.---} For three-level atoms ($\Omega_c\neq0$), the photon-mediated dipole-dipole interactions described above lead to coupling also between Rydberg pseudo-spins $\hat{\sigma}_n=|g\rangle_n\langle r|$ of the array atoms $n=1,...,N$. This is revealed by a formulation of collective Rydberg surface modes of the array, defined by
\begin{equation}
\hat{\sigma}_{\mathbf{k}}=\frac{1}{\sqrt{N}}\sum_{n=1}^N e^{-i \mathbf{k}\cdot \mathbf{r}_n}\hat{\sigma}_n,
\quad
\hat{\sigma}_n=|g\rangle_n\langle r|,
\label{sig}
\end{equation}
where $\mathbf{r}_n$ is the lattice position of atom $n$. To this end, we start with the full Hamiltonian of the array atoms and quantized free-space photon modes, assuming a coherent-state input field on the $|g\rangle\leftrightarrow |e\rangle$ transition with an average amplitude $\sum_{\mathbf{k}}e^{i\mathbf{k}\cdot\mathbf{r}_n}E_{p,\mathbf{k}}$ (continuous wave, frequency $\omega_p$) on an atom $n$, in addition to the uniform coupling field $\Omega_c$. Van der Waals (vdW) interaction potential $V_{nm}$ between pairs of atoms ($n$ and $m$) populating Rydberg levels is also considered. We then derive a Heisenberg-Langevin equation for $\hat{\sigma}_{\mathbf{k}}$ using the following \cite{SI}: (i) all photon modes are eliminated (Born-Markov approximation), (ii) assuming weak field, internal state saturation is neglected (linearization in $E_{p,\mathbf{k}}$), (iii) the states $|e\rangle$ are adiabatically eliminated due to their fast cooperative decay $\Gamma_{\mathbf{k}}$, (iv) decay of the long-lived Rydberg level is neglected. Finally, we obtain (in a laser-rotated frame) \cite{SI}
\begin{eqnarray}
\dot{\hat{\sigma}}_{\mathbf{k}}&=&i\left[\sum_{\mathbf{k}}\hat{H}^{\mathrm{eff}}_{\mathbf{k}}+\hat{V},\hat{\sigma}_{\mathbf{k}}\right]-\hat{F}_{\mathbf{k}}(t),
\label{HL}
\\
\hat{H}^{\mathrm{eff}}_{\mathbf{k}}&=&\left(\omega_{\mathbf{k}}-\delta-i\frac{\gamma_{\mathbf{k}}}{2}\right)\hat{\sigma}_{\mathbf{k}}^{\dag}\hat{\sigma}_{\mathbf{k}}
-\left(i\Omega_{\mathbf{k}}\hat{\sigma}_{\mathbf{k}}^{\dag}+\mathrm{h.c.}\right).
\label{H}
\end{eqnarray}
The effective Hamiltonian $\hat{H}^{\mathrm{eff}}_{\mathbf{k}}$ (non-Hermitian), together with the linearized commutation relations $[\hat{\sigma}_{\mathbf{k}},\hat{\sigma}_{\mathbf{k}'}^{\dag}]\approx \delta_{\mathbf{k},\mathbf{k}'}$, does not mix Rydberg surface modes of different momenta $\mathbf{k}$. Each mode acquires a collective shift $\omega_{\mathbf{k}}$ and width $\gamma_{\mathbf{k}}$ with respect to the ``bare" two-photon transition resonance (detuning) $\delta=\delta_p+\delta_c$,
\begin{eqnarray}
i\omega_{\mathbf{k}}+\frac{\gamma_{\mathbf{k}}}{2}=-\frac{2|\Omega_c|^2}{\Gamma_{\mathbf{k}}}r_{\mathbf{k}},
\;\
r_{\mathbf{k}}=-\frac{\Gamma_{\mathbf{k}}}{\Gamma_{\mathbf{k}}+\gamma_{\mathrm{sc}}+i2(\Delta_{\mathbf{k}}-\delta_p)}.
\label{del}
\end{eqnarray}
Here $r_{\mathbf{k}}$ is the reflectivity of a two-level atom array \cite{coop} and $\gamma_{\mathrm{sc}}\ll \Gamma_{\mathbf{k}}$ accounts for scattering losses (see below).
Equation (\ref{del}) is the array collective generalization of the two-photon transition resonance in an individual three-level atom \cite{PET}: the cooperative $|g\rangle\rightarrow |e\rangle$ decay $\Gamma_{\mathbf{k}}$ replaces that of an individual atom $\gamma_{\mathrm{atom}}$ and the corresponding cooperative shift $\Delta_{\mathbf{k}}$ is included. In addition, each collective Rydberg mode $\hat{\sigma}_{\mathbf{k}}$
is driven by a two-photon field at the corresponding direction $\mathbf{k}$, containing an average coherent drive $\Omega_{\mathbf{k}}$ and a quantum Langevin noise $\hat{F}_{\mathbf{k}}$,
\begin{eqnarray}
\Omega_{\mathbf{k}}+\hat{F}_{\mathbf{k}}(t)=-r_{\mathbf{k}}\frac{2\Omega_c}{\Gamma_{\mathbf{k}}}
\left(\sum_{s=\pm}\hat{E}_{\mathrm{in},\mathbf{k},s}\frac{\sqrt{N}d^{\ast}}{\hbar}\right).
\label{Om}
\end{eqnarray}
Here $\hat{E}_{\mathrm{in},\mathbf{k},s}$ is the input field on the $|g\rangle\leftrightarrow |e\rangle$ transition (dipole matrix element $d$), either right- ($s=+$) or left- ($s=-$) propagating, containing the average coherent amplitude $E_{p,\mathbf{k}}\delta_{s,+}$ (right-propagating $s=+$) and a vacuum-field Langevin noise $\hat{E}_{0,\mathbf{k},s}(t)$.
Finally, the only term in Eq. (\ref{HL}) that can mix the normal modes $\mathbf{k}$ is the nonlinear interaction $\hat{V}=(1/2)\sum_{n\neq m} V_{nm}\hat{\sigma}_n^{\dag}\hat{\sigma}_n \hat{\sigma}_m^{\dag}\hat{\sigma}_m$. However, in  the blockade regime considered below, directionality $\mathbf{k}$ is conserved even in the nonlinear regime.

\begin{figure}[t]
  \begin{center}
    \includegraphics[width=\columnwidth]{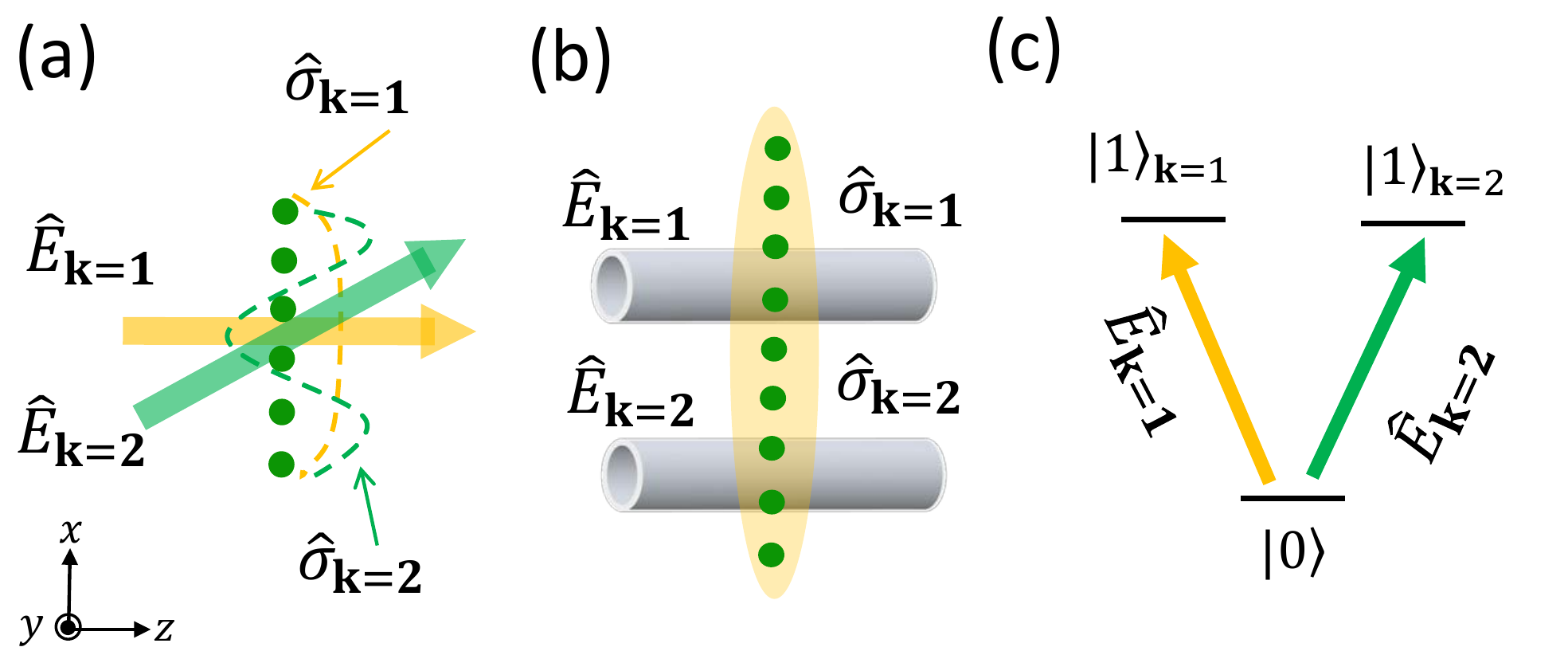}
    \caption{\small{
Effective multi-channel waveguide QED model. Each field component $\mathbf{k}$ is directionally coupled to a corresponding Rydberg surface mode $\hat{\sigma}_{\mathbf{k}}$ (a), in analogy to multiple 1D waveguide channels with index $\mathbf{k}$ (b). (c) Rydberg blockade restricts the relevant atomic levels to include singly-excited collective modes, $|1\rangle_{\mathbf{k}}=\sigma_{\mathbf{k}}^{\dag}|0\rangle$ coupled by the field component $\mathbf{k}$ to a common ground state $|0\rangle$.
    }} \label{fig2}
  \end{center}
\end{figure}

\emph{Directional scattering.---}
The directional light-matter coupling is revealed also in the output light, whose $\mathbf{k}$ component is found to be (rotated at $\omega_p$) \cite{SI},
\begin{equation}
\hat{E}_{\mathbf{k},\pm}(z)=e^{\pm i k_z z}\left[(1+r_{\mathbf{k}})\hat{E}_{\mathrm{in},\mathbf{k},\pm}+r_{\mathbf{k}}\hat{E}_{\mathrm{in},\mathbf{k},\mp}
+r_{\mathbf{k}}\frac{\hbar\Omega^{\ast}_c}{d\sqrt{N}}\hat{\sigma}_{\mathbf{k}}\right].
\label{E}
\end{equation}
Here $\pm$ denotes right- (left-) propagating field at $z>0$ ($z<0$),
$\hat{E}_{\mathrm{in},\mathbf{k},\pm}$ is the input field defined above, $k_z=\sqrt{(\omega_p/c)^2-|\mathbf{k}|^2}$, and paraxial light $|\mathbf{k}|\ll \omega_p/c$ is assumed. For $\Omega_c=0$, only the first two terms exist and we obtain the directional reflection of the input quantum field known for a two-level atom array \cite{coop}. Ideally, for $\gamma_{\mathrm{sc}}=0$ , this yields perfect reflection $r_{\mathbf{k}}= -1$ at the collective single-photon resonance $\delta_p=\Delta_{\mathbf{k}}$. Accounting for non-directional scattering losses due to disorder in the array yields a finite $\gamma_{\mathrm{sc}}\ll \Gamma_{\mathbf{k}}$ (see below) and an on-resonance reflectivity magnitude $r_{\mathrm{res},\mathbf{k}}=\Gamma_{\mathbf{k}}/(\Gamma_{\mathbf{k}}+\gamma_{\mathrm{sc}})$. Thus, $r_{\mathrm{res},\mathbf{k}}<1$  characterizes the strength of directional light-matter coupling of the array, in analogy to the figure of merit of waveguide QED, $\Gamma_{\mathrm{1d}}/(\Gamma_{\mathrm{1d}}+\gamma_{\mathrm{sc}})$ (Fig. 1a).

When the coupling field is turned on ($\Omega_c \neq 0$), the corresponding collective Rydberg excitation $\hat{\sigma}_{\mathbf{k}}$ adds a contribution to the scattering. Without interactions, $\hat{V}=0$, and by solving for the steady state of $\hat{\sigma}_{\mathbf{k}}$, we obtain for $\hat{E}_{\mathbf{k}}(z)$ an expression similar to the first two terms of Eq. (\ref{E}), however with a modified reflectivity $\tilde{r}_{\mathbf{k}}$,
\begin{eqnarray}
\tilde{r}_{\mathbf{k}}=r_{\mathbf{k}}\frac{\delta}{\delta-\omega_{\mathbf{k}}+i\gamma_{\mathbf{k}}/2}.
\label{rt}
\end{eqnarray}
That is, for a large two-photon detuning $|\delta|\gg |\gamma_{\mathbf{k}}/2+i\omega_{\mathbf{k}}|$ we obtain the two-level reflectivity $\tilde{r}_{\mathbf{k}}\rightarrow r_{\mathbf{k}}$, whereas close to bare two-photon resonance $|\delta|\ll |\gamma_{\mathbf{k}}/2+i\omega_{\mathbf{k}}|$ the reflectivity tends to zero $\tilde{r}_{\mathbf{k}}\rightarrow 0$ and light is transmitted. This is the essence of electromagnetically induced transparency (EIT) in the 2D array: tuning $\delta$ to switch from strong reflection to strong transmission, keeping directionality in both cases (see also \cite{RIV,POH4}). Close to collective single-photon resonance $|\delta_p-\Delta_{\mathbf{k}}|\ll \Gamma_{\mathbf{k}}$ this defines $\gamma_{\mathbf{k}}\approx \gamma^{(0)}_{\mathbf{k}}\equiv4|\Omega_c|^2/\Gamma_{\mathbf{k}}$ as the EIT transparency width, which is the collective 2D array version of the single-atom transparency width $\sim 4|\Omega_c|^2/\gamma_{\mathrm{atom}}$.

\emph{Nonlinearity.---}
Without the interaction $\hat{V}$ the excitations of the essentially bosonic (linearized) collective modes $\hat{\sigma}_{\mathbf{k}}$ exhibit a harmonic spectrum with spacings $\omega_{\mathbf{k}}$ [Hamiltonian (\ref{H})]. The vdW interaction $V_{nm}=V(|\mathbf{r}_n-\mathbf{r}_m|)$ adds a nonlinearity that shifts the energy of two excitations by at least $\sim \mathrm{min} V_{nm}$. When this shift is much larger than the collective width $\gamma_{\mathbf{k}}$, two-excitation states of the Rydberg modes are unattainable. This Rydberg blockade regime is reached when the cross-section of the incident beam is smaller than the blockade radius $R_b$ defined by $V(R_b)=|\gamma_{\mathbf{k}}+i2( \omega_{\mathbf{k}}-\delta)|\sim \gamma_{\mathbf{k}}$, as we show explicitly in \cite{SI} for the case of paraxial light.

\emph{Effective waveguide QED model.---}
The above considerations establish the following effective model (Fig. 2). A field component $\hat{E}_{\mathbf{k}}(z)$ at direction (in-plane wavevector) $\mathbf{k}$ is exclusively coupled to a corresponding Rydberg surface ``dipole" $\hat{\sigma}_{\mathbf{k}}$ at longitudinal position $z=0$, resulting in $\mathbf{k}$-preserving scattering (Fig. 2a). This is analogous to scattering in a 1D waveguide channel, and generalizes to multiple 1D ``channels" $\mathbf{k}$ when several components $\hat{E}_{\mathbf{k}}(z)$ exist at the input (Fig. 2b). Rydberg blockade restricts the Hilbert space to include only singly-excited states of the surface modes $\mathbf{k}$, $|1\rangle_{\mathbf{k}}=\hat{\sigma}_{\mathbf{k}}^{\dag}|0\rangle$, which share a common ground state $|0\rangle=|gg....g\rangle$ (of all atoms), coupled to them by the corresponding fields  $\hat{E}_{\mathbf{k}}(z)$ (Fig. 2c). This generates interactions between photons of different channels $\mathbf{k}$, where the excitation of one channel from the common ground state prevents the excitation of all other channels. Importantly, this interaction conserves directionality $\mathbf{k}$ since no excitations are exchanged between channels.
Calculations of any physical observables are carried out by first evaluating atomic observables using the effective Hamiltonian $\sum_{\mathbf{k}}\hat{H}^{\mathrm{eff}}_{\mathbf{k}}$ from Eq. (\ref{H}) with the replacement $\hat{\sigma}_{\mathbf{k}}\rightarrow |0\rangle \langle 1|_{\mathbf{k}}$ (including quantum jump/noise terms corresponding to $\gamma_{\mathbf{k}}$), then relating them to field observables via Eq. (\ref{E}) \cite{SI}.

\begin{figure}[t]
\includegraphics[width=\columnwidth]{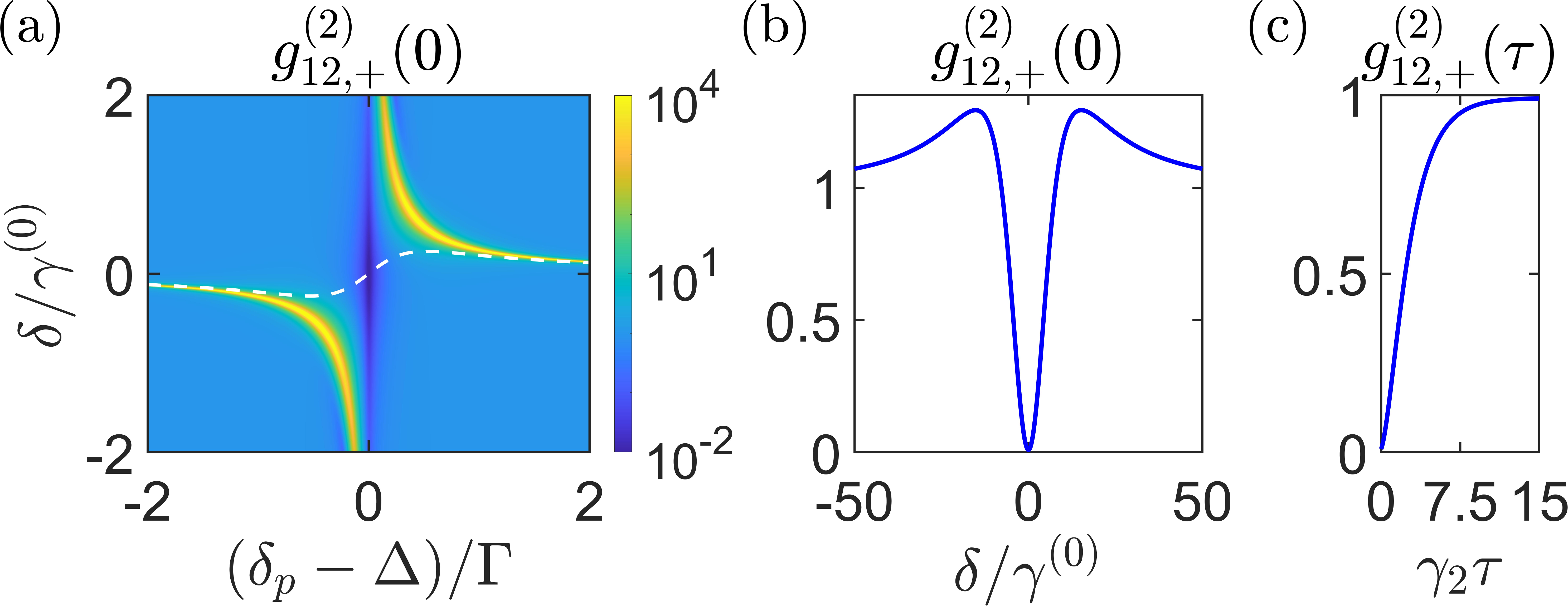}
\caption{\small{
Inter-channel photon correlations in transmission [$s=+$ in Eqs. (\ref{G1}), (\ref{G2})]. (a) Cross-bunching $g^{(2)}_{\mathbf{12}}(0)>1$ is strongest along the hyperbole of full reflection (see text), whereas cross-antibunching $g^{(2)}_{\mathbf{12}}(0)<1$ is prominent along single-photon resonance $\delta_p=\Delta$. (b) Projection of (a) along $\delta_p=\Delta$ (with extended range). (c) Correlation time is set by the collective width $\gamma_{\mathbf{k}=\mathbf{2}}$. Parameters: $\gamma_{\mathrm{sc}}/\Gamma=0.05$ in all plots, $\delta=0$ in (c).
}}
\end{figure}

\emph{Multi-channel correlations and entanglement.---}
Considering the excitation of two channels $\mathbf{k}= \mathbf{1},\mathbf{2}$ with corresponding coherent input fields $E_{p,\mathbf{1}}$ and $E_{p,\mathbf{2}}$, we calculate the steady-state intensity $G^{(1)}_{\mathbf{k},s}=\langle \hat{E}^{\dag}_{\mathbf{k},s}(t)\hat{E}_{\mathbf{k},s}(t)\rangle$ of the output of channel $\mathbf{k}=\mathbf{1}$
in transmission ($s=+$) and reflection ($s=-$), obtaining \cite{SI}
\begin{eqnarray}
&&\frac{G^{(1)}_{\mathbf{1},s}}{|E_{p,\mathbf{1}}|^2}=|b_s|^2+\frac{|r-\tilde{r}|^2-\left[b_s^{\ast}(r-\tilde{r})+\mathrm{c.c.}\right]}{1+S_{\mathbf{2}}}.
\label{G1}
\end{eqnarray}
Here $S_{\mathbf{k}}=\frac{2|\Omega_{\mathbf{k}}|^2}{(\omega_{\mathbf{k}}-\delta)^2+(\gamma_{\mathbf{k}}/2)^2}$, $b_+=(1+r)$, $b_-=r$, and weak field is assumed for channel $\mathbf{k}= \mathbf{1}$, $S_{\mathbf{1}}\ll 1$.
For simplicity of presentation, we took identical parameters $\Delta_{\mathbf{k}}$ and $\Gamma_{\mathbf{k}}$
for both channels, so that the index $\mathbf{k}$ in $\alpha_{\mathbf{k}}$, with $\alpha\in\{\Delta,\Gamma,\omega,\gamma,r,\tilde{r}\}$,
is dropped unless specified otherwise (see \cite{SI} for full results).
We observe that strong field in one channel, $S_{\mathbf{2}}\gg 1$, can convert transmission into reflection in the other channel, $G^{(1)}_{\mathbf{1},+}\rightarrow0$, due to the Rydberg interaction. This ``cross-blockade" yields cross correlations between the channels which can be evaluated by the second-order coherence
$G^{(2)}_{\mathbf{k}\mathbf{k}',\pm}(\tau)=\langle \hat{E}^{\dag}_{\mathbf{k},\pm}(t)\hat{E}^{\dag}_{\mathbf{k}',\pm}(t+\tau)\hat{E}_{\mathbf{k}',\pm}(t+\tau)\hat{E}_{\mathbf{k},\pm}(t)\rangle$ in steady state. For zero-delay $\tau=0$ and weak fields $S_{\mathbf{1}},S_{\mathbf{2}}\ll 1$ we obtain \cite{SI}
\begin{eqnarray}
\frac{G^{(2)}_{\mathbf{1}\mathbf{2},s}(0)}{|E_{p,\mathbf{1}}|^2|E_{p,\mathbf{2}}|^2}&=&
|b_s|^2\left|b_s-2(r-\tilde{r})\right|^2.
\label{G2}
\end{eqnarray}
In Fig. 3a we plot the photon cross-correlation $g^{(2)}_{\mathbf{1}\mathbf{2},s}(\tau)=G^{(2)}_{\mathbf{1}\mathbf{2},s}(\tau)/[G^{(1)}_{\mathbf{1},s}G^{(1)}_{\mathbf{2},s}]$ for the transmitted field at zero delay, revealing features of both ``cross-antibunching" $g^{(2)}_{\mathbf{1}\mathbf{2},+}(0)<1$ and ``cross-bunching" $g^{(2)}_{\mathbf{1}\mathbf{2},+}(0)>1$. The former are mostly evident at single-photon collective resonance, i.e. along the vertical line $\delta_p=\Delta$, as also depicted in Fig. 3b. These cross-antibunching correlations are strongest near the bare two-photon resonance $\delta\ll \gamma^{(0)}=4|\Omega_c|^2/\Gamma$, where EIT allows single-photon transmission while the blockade inhibits that of two-photons, obtaining $g^{(2)}_{\mathbf{1}\mathbf{2},+}(0)\approx |1-r^2|^2\ll 1$.
For input light only at a single channel $\mathbf{k}=\mathbf{1}$ we get the same expressions with the replacement $\mathbf{2}\rightarrow \mathbf{1}$ yielding single-mode antibunching also observed in Ref. \cite{POH4} (and in analogy to Ref. \cite{CHA4} and standard waveguide QED \cite{CHA}).
Cross-channel bunching is most prominently seen in Fig. 3a along the hyperbole $\frac{\delta}{\gamma^{(0)}}=\frac{\Gamma/4}{\delta_p-\Delta}$, for which the linear reflection, Eq. (\ref{rt}), is maximal (complete absorption), $\tilde{r}=-r_{\mathrm{res}}=-\Gamma/(\Gamma+\gamma_{\mathrm{sc}})$ \cite{SI}. In this case, linear transmission is inhibited but Rydberg blockade allows for transmission of bunched multi-channel photon pairs, $g^{(2)}_{\mathbf{1}\mathbf{2},+}(0)=\frac{|1+r|^2|1-r-2r_{\mathrm{res}}|^2}{|1-r_{\mathrm{res}}|^4}>1$.
In reflection ($s=-$) bunching is exhibited for small $\delta$, where $g^{(2)}_{\mathbf{1}\mathbf{2},-}(0)\approx 1/|\delta|^4\gg 1$.
We also obtain analytically the time delayed correlations $g^{(2)}_{\mathbf{1}\mathbf{2},s}(\tau)$ without assuming identical channel parameters \cite{SI}. These exhibit exponentials with the collective width $\gamma_{\mathbf{k}=\mathbf{2}}$ corresponding to the channel of the subsequently detected photon ($\mathbf{k}=\mathbf{2}$). As seen in Fig. 3c, this sets the correlation time $1/\gamma_{\mathbf{2}}$, as expected from the collective mode picture of Hamiltonian (\ref{H}).

It is particulary interesting to study the physics near two-photon resonance of the collective Rydberg modes, $\delta=\omega=-(\gamma^{(0)}/2)\mathrm{Im}[r]$. This resonance condition yields a relation between $\delta$ and $(\delta_p-\Delta)$ shown by the dashed curve in Fig. 3a: the dispersive shape follows from the imaginary part of $r$ in Eq. (\ref{del}). In Fig. 4a we plot $g^{(2)}_{\mathbf{12}}$ along this curve, observing strong cross-correlations even at very large single-photon detunings,
where they saturate to $g^{(2)}_{\mathbf{12},+}\rightarrow (1-2r_{\mathrm{res}})^2/(1-r_{\mathrm{res}})^4\gg 1$. Therefore, inter-channel correlations can become strong at collective two-photon Rydberg resonance $\delta=\omega$.

\begin{figure}[h]
    \includegraphics[width=\columnwidth]{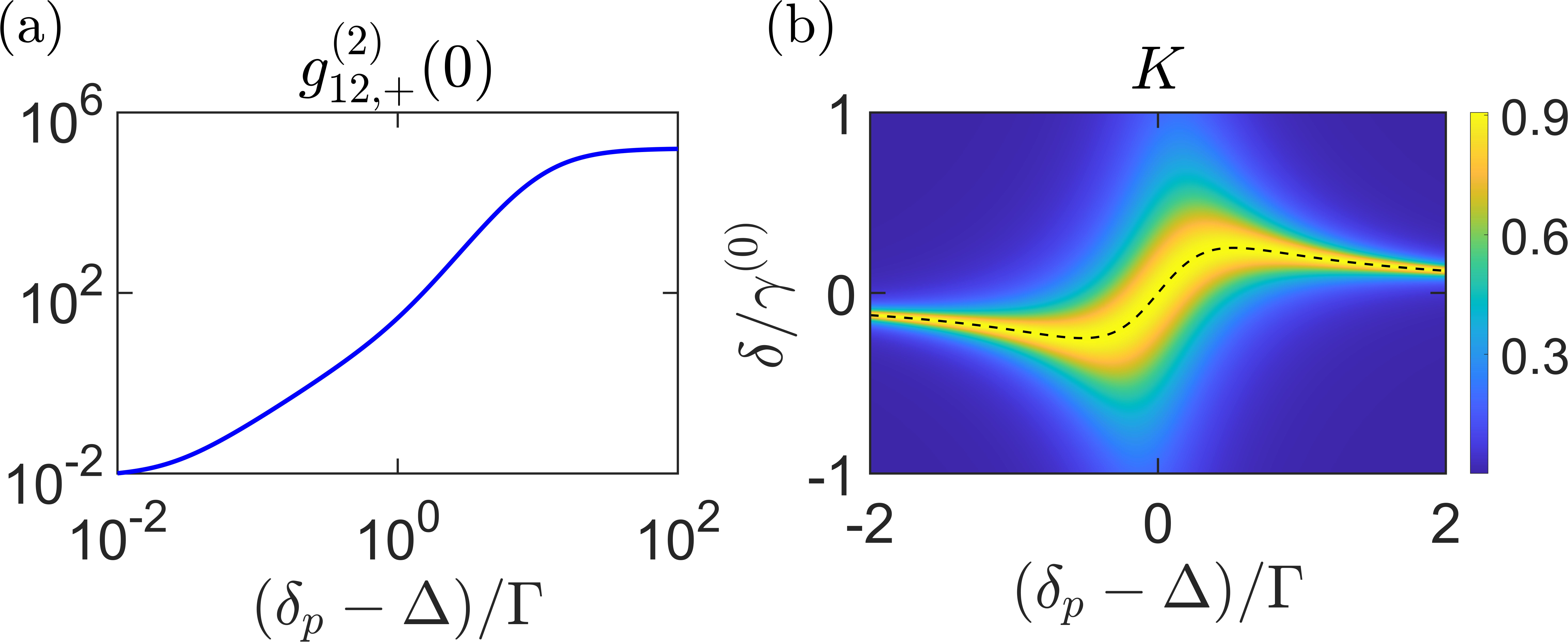}
    \caption{\small{
    Correlations and entanglement near collective two-photon Rydberg resonance $\delta=\omega$. (a) Projection of Fig. 3a along the curve $\delta=\omega$ (dashed line in Fig. 3a). (b) Inter-channel quantum correlations, $K$ from Eq. (\ref{D}), peak along the curve $\delta=\omega\propto \mathrm{Im}[r]$ (dashed line) with a width $\gamma\propto \mathrm{Re}[r]$. Parameters: $\gamma_{\mathrm{sc}}/\Gamma=0.05$ (yielding $r^2_{\mathrm{res}}\approx0.91$).
    }}
\end{figure}

To examine whether these inter-channel correlations are quantum, we calculate the Duan \emph{et al.} inseparability parameter $D$ \cite{DUA} between the mode operators $\hat{E}_{\mathbf{1},s}(z)$ and $\hat{E}_{\mathbf{2},s}(z)$, where $D<1$ signifies an entangled state \cite{SI}. We also quantify quantum correlations by evaluating the quadrature variance $\mathcal{V}$ of the total field $\hat{E}_{\mathbf{1},s}+\hat{E}_{\mathbf{2},s}$, where $\mathcal{V}<1$ signifies two-mode quantum squeezing and correlations. For weak input fields with $E_{p,\mathbf{2}}=e^{-i\varphi}E_{p,\mathbf{1}}$ we find both in transmission and reflection \cite{SI}
\begin{eqnarray}
D=\mathcal{V}=1-2\left|\frac{E_{p,\mathbf{1}}}{E_0}\right|^2 K, \;\; K=\frac{r_{\mathrm{res}}^2}{1+4(\delta-\omega)^2/\gamma^2}.
\label{D}
\end{eqnarray}
In general, $\mathcal{V}$ depends sinusoidally on the phase difference $\varphi$ \cite{SI}, whereas here we took $\varphi=\pi/2$ for simplicity,
and the expression is valid for weak fields, where $2|E_{p,\mathbf{1}}/E_0|^2\ll 1$ and $E_0$ is the zero-point field of a mode $\hat{E}_{\mathbf{k},s}(z)$ \cite{SI}. The normalized, mode-independent quantity, $K>0$, signifies entanglement ($D<1$) and quantifies two-mode quantum correlations (squeezing $\mathcal{V}<1$) between the two channels $\mathbf{k}=\mathbf{1},\mathbf{2}$. It appears as a Lorentzian peaked at the collective Rydberg resonance $\delta=\omega$. Therefore, inter-channel entanglement, as characterized by $D$ and $\mathcal{V}$, is optimal when the collective Rydberg modes are excited resonantly, and not necessarily at bare two-photon ($\delta=0$) and collective single-photon ($\delta_p=\Delta$) resonances, see Fig 4b.  This result matches the physical picture provided by the effective model of Fig. 2 and Eq. (\ref{H}); namely, the collective Rydberg excitations are the relevant nonlinear atomic system that can entangle photons. Importantly, although the Rydberg resonance $\delta=\omega$ does not in general coincide with the single-photon resonance $\delta_p=\Delta$, the peak value of $K$ in Eq. (\ref{D}) is nevertheless set by the resonant two-level atom-array reflectivity $r_{\mathrm{res}}=\Gamma/(\Gamma+\gamma_{\mathrm{sc}})$. This establishes $r_{\mathrm{res}}$ as a figure of merit for multi-channel quantum correlations at various regimes.

\emph{Practical considerations.---}
The above description, formulated for an infinite array and incident plane waves $\mathbf{k}$, provides an excellent approximation for realistic finite-size incident Gaussian beams directed at corresponding ``angles" $\mathbf{k}$, and with beam-waist smaller than the array size, $w<\sqrt{N}a$ \cite{coop,note1}. Directionality $\mathbf{k}$ is conserved as long as the interaction potential is effectively uniform across the beam, requiring $w<R_b$.
Here $R_b=(C_6/\gamma_{\mathbf{k}})^{1/6}$ is the blockade radius defined above, with $C_6$ the realization-dependent coefficient of the vdW interaction $V(r)=C_6/r^6$. For example, considering an array realized by a 2D optical lattice of $^{87}$Rb atoms as in \cite{RUI}, with $a/\lambda=532/780$ and a Rydberg level with principal quantum number $n=100$ and $\Omega_c=2\pi\times 0.75$ MHz, we estimate $R_b\sim 15$ $\mu$m \cite{SI}.
Considering weak disorder in the array positions, e.g. due to atomic motion, we find, using the approach of Refs. \cite{AAMO,om2}, that the non-directional scattering rate discussed above scales as $\gamma_{\mathrm{sc}}\sim \eta^2\Gamma_{\mathbf{k}}$, where $\eta\sim10^{-1}\ll 1$ is the Lamb-Dicke parameter of the atomic traps.

\emph{Discussion.---}
This work studies the quantum optics of 2D arrays of Rydberg atoms by establishing its mapping to a multi-channel model of waveguide QED. Nonlinear scattering is revealed to be dominated by collective Rydberg resonances of the array, generating inter-channel quantum correlations whose strength is characterized by the array reflectivity. These results suggest that atom arrays are naturally suited to study novel multimode quantum nonlinear optics, in the context of quantum information and many-body physics, going beyond typical waveguide QED regimes. For example, the inter-channel correlations discussed above motivate the study of multimode many-photon bound states which generalize those known in 1D \cite{FAN,BAR}, or applications in quantum gates involving multiple transverse modes \cite{DK,BARg,RIV}. In particular, the analysis of such effects near collective two-photon Rydberg resonance would be important, and the consideration of multiple 2D arrays should open new possibilities. Remarkably, the atom-array realization allows to explore these prospects using free-space fields scattered off several dozen atoms.


\begin{acknowledgments}
We acknowledge fruitful discussions with Darrick Chang and Ofer Firstenberg, and financial support from the Israel Science Foundation (grant No. 2258/20), the Center for New Scientists at the Weizmann Institute of Science, and the Council for Higher Education (Israel). This research is made possible in part by the historic generosity of the Harold Perlman Family.
\end{acknowledgments}

\newpage\null\newpage
\widetext
\begin{center}
    {\textbf{Multi-channel waveguide QED with atomic arrays in free space:\\Supplementary Material}}
\end{center}
\setcounter{equation}{0}
\setcounter{figure}{0}
\setcounter{table}{0}
\setcounter{page}{1}
\makeatletter
\renewcommand{\theequation}{S\arabic{equation}}
\renewcommand{\thefigure}{S\arabic{figure}}
\renewcommand{\bibnumfmt}[1]{[S#1]}
\renewcommand{\citenumfont}[1]{S#1}
\section{1. General formalism}
\subsection{1.1. Hamiltonian and Heisenberg-Langevin equations}
We consider a 2D array of identical atoms $n=1,...,N$ spanning the $xy$ plane at positions $\mathbf{r}_n$ (at $z=0$ $\forall n$) and forming an infinite square lattice with lattice constant $a$. Each atom is a three-level ladder system as shown in figure 1c in the main text, composed of levels $\ket{g}$, $\ket{e}$, $\ket{r}$, where $\ket{r}$ is a metastable Rydberg level. Pairs of atoms $n$ and $m$ which populate Rydberg levels interact via the van der Waals (vdW) potential $V(\textbf{r}_{nm})$ with $\textbf{r}_{nm}=\textbf{r}_{n}-\textbf{r}_{m}$. In addition, the atoms are illuminated by coherent-state fields on both transitions, which can be represented by ``classical" (external) driving fields + a quantum field in an initial vacuum state. The total Hamiltonian is given by
\begin{equation}\label{Seq1}
    \hat{H}=\hat{H}_f+\hat{H}_s+\hat{H}_I.
\end{equation}
Here  $\hat{H}_f=\sum_{\mathbf{k}}\sum_{k_z}\sum_{\mu}\hbar\nu_{\mathbf{k}k_z}\hat{a}^{\dagger}_{\textbf{k}k_z\mu}\hat{a}_{\textbf{k}k_z\mu}$ is the Hamiltonian of the photon modes in free space, characterized by the in-plane and longitudinal wavevectors $\mathbf{k}=(k_x,k_y)$ and $k_z$, respectively, and the polarization $\mu$, with $\nu_{\mathbf{k}k_z}=c\sqrt{|\mathbf{k}|^2+k_z^2}$. The Hamiltonian of the atomic system is given by
\begin{equation}
\hat{H}_s=\sum_n\bigg[\hbar\omega_{e}\hat{\sigma}_{ee,n}+\hbar\omega_{r}\hat{\sigma}_{rr,n}\bigg]+\sum_{n>m}V(\textbf{r}_{nm})\hat{\sigma}_{rr,n}\hat{\sigma}_{rr,m}-\sum_n\bigg[\hbar\Omega_p(\textbf{r}_n)e^{-i\omega_pt}\hat{\sigma}_{ge,n}^\dagger+\hbar\Omega_ce^{-i\omega_ct}\hat{\sigma}_{er,n}^\dagger+\mathrm{h.c}\bigg].
\end{equation}
where $\hat{\sigma}_{\alpha\alpha',n}=|\alpha\rangle_n\langle \alpha'|$ for an atom $n$ and $\hbar\omega_{\alpha}$ is the energy of level $|\alpha\rangle$. Here $\Omega_p(\mathbf{r}_n)=|\mathbf{d}_{ge}| E_p(\mathbf{r}_n)/\hbar$ is the Rabi frequency due to the coherent input field $E_p(\mathbf{r}_n)$ at frequency $\omega_p$ coupled to the transition $\ket{g}\leftrightarrow \ket{e}$ (dipole matrix element $\mathbf{d}_{ge}$) of the atom $n$ , whereas $\Omega_c$ is the Rabi frequency of the uniform average field at frequency $\omega_c$ working on the  $\ket{e}\leftrightarrow \ket{r}$ transition of the atoms.

The interaction Hamiltonian between the quantum field and the atoms is given by
\begin{equation}
\hat{H}_I=-\sum_n\left[\textbf{d}_{ge} \hat{\sigma}_{ge,n}+\textbf{d}_{er} \hat{\sigma}_{er,n}+\mathrm{h.c.}\right]\cdot
\left[ \hat{\mathbf{E}}(\mathbf{r}_n)+ \hat{\mathbf{E}}^{\dag}(\mathbf{r}_n)\right]
\label{SHI}
\end{equation}
where $\mathbf{d}_{\alpha\alpha'}$ is the dipole matrix element corresponding to the $\ket{\alpha}\leftrightarrow \ket{\alpha'}$ transition and the quantum field operator is given by
\begin{equation}
    \hat{\mathbf{E}}(\textbf{r})=i\sum_{\mathbf{k}}\sum_{k_z}\sum_{\mu}\sqrt{\frac{\hbar\nu_{\mathbf{k}k_z}}{2\epsilon_0V}}\mathbf{e}_{\textbf{k}k_z\mu}\hat{a}_{\textbf{k}k_z\mu}e^{i(\textbf{k}\cdot\textbf{r}+k_zz)},
\label{SE}
\end{equation}
with $V$ the quantization volume and $\mathbf{e}_{\mathbf{k}k_z\mu}$ the photon polarization vector.

We derive Heisenberg-Langevin equations for the atomic operators in the laser-rotated frame, by integrating out the field operators within the Born-Markov approximation (assuming also the frequencies $\omega_p$, $\omega_c$ and their difference represent fast time scales). By further assuming that all atoms are initially in the ground state $|g\rangle$ and that the field $\Omega_p$ is weak, we linearize the equations by keeping terms up to first order in $\Omega_p$, with the exception of the strongly interacting Rydberg vdW potential term. The resulting equations for the relevant laser-frame operators, $\tilde{\sigma}_{ge,n}=e^{i\omega_p t} \hat{\sigma}_{ge,n}$ and $\tilde{\sigma}_{gr,n}=e^{i(\omega_p+\omega_c) t} \hat{\sigma}_{gr,n}$ are given by
\begin{equation}
    \begin{split}
&\frac{\partial\tilde{\sigma}_{ge,n}}{\partial t}=\bigg(i\delta_p-\frac{\gamma_{\mathrm{sc}}}{2}\bigg)\tilde{\sigma}_{ge,n}+\frac{i}{\hbar}\frac{\omega_p^2}{c^2\epsilon_0}\sum_{m}\textbf{d}^*_{ge}\cdot\textbf{G}(\omega_p,\textbf{r}_{nm})\cdot\textbf{d}_{ge}\tilde{\sigma}_{ge,m}+i\Omega_p(\textbf{r}_n)+i\Omega_c^*\tilde{\sigma}_{gr,n}+\hat{F}_{ge}(\textbf{r}_n),\\
&\frac{\partial\tilde{\sigma}_{gr,n}}{\partial t}=i\delta\tilde{\sigma}_{gr,n}+i\Omega_c\tilde{\sigma}_{ge,n}-\frac{i}{\hbar}\sum_{m\neq n}V(\textbf{r}_{nm})\tilde{\sigma}_{gr,m}^\dagger\tilde{\sigma}_{gr,m}\tilde{\sigma}_{gr,n},\\
    \end{split}
\label{SEOM1}
\end{equation}
where $\delta_p=\omega_p-\omega_e$ is the detuning of the $\ket{g}\leftrightarrow \ket{e}$ transition and $\delta=\delta_p+\delta_c$ is the two-photon detuning, with $\delta_c=\omega_c-(\omega_r-\omega_e)$ being the detuning of the $\ket{e}\leftrightarrow \ket{r}$ transition. Here $\textbf{G}(\omega,\textbf{r})$ is the (tensor) Green's function of the photon field in free space at frequency $\omega$ which depends only on the distance difference $\mathbf{r}$ (translation invariance) \cite{Scoop,SChapterOne}. This Green's function gives rise to collective radiation and photon-mediated dipole-dipole interaction between the transitions $\ket{g}\leftrightarrow \ket{e}$ of different atoms. We add $\gamma_{\mathrm{sc}}$ to the equations to account for scattering losses due to weak disorder in the array positions \cite{SChapterOne,Smotion}.\\
The corresponding quantum Langevin noise term is given by $\hat{F}_{ge}(\textbf{r}_n,t)=i\frac{\textbf{d}^{\ast}_{ge}\cdot\hat{\mathbf{E}}_0(\textbf{r}_n,t)}{\hbar}$, where $\hat{\mathbf{E}}_0(\textbf{r}_n,t)$ is the input quantum vacuum field (noise of coherent input) given by Eq. (\ref{SE}) with the replacement $\hat{a}_{\mathbf{k}k_z\mu}(t)\rightarrow \hat{a}_{\mathbf{k}k_z\mu}(0) e^{-i(\nu_{\mathbf{k}k_z}-\omega_p)t}$. In principle, $\gamma_{\mathrm{sc}}$ adds a correction to the Langevin noise, however, this correction is negligible and irrelevant for our purposes. 
We note that the dipole-dipole interaction between the transitions $\ket{e} \leftrightarrow \ket{r}$ of different atoms is of higher order in $\Omega_p$ and disappears from the equations after linearization. Furthermore, a decay term $-\gamma_r\tilde{\sigma}_{gr,n}$ (and corresponding quantum noise) in the second equation, due to the radiative lifetime of the long-lived Rydberg level $r$, is neglected here assuming $\gamma_r^{-1}$ to be a long timescale with respect to relevant dynamics.

\subsection{1.2.  Transformation to momentum space}
The lattice translation invariance of the array motivates the transformation of the equations to a basis of lattice momentum $\mathbf{k}=(k_x,k_y)$ with $k_{x,y}\in[-\pi/a,\pi/a]$ within the Brillouin zone. The transformation is defined via
\begin{equation}
\begin{split}
        &\tilde{\sigma}_{\alpha\alpha',\textbf{k}}=\frac{1}{\sqrt{N}}\sum_n\tilde{\sigma}_{\alpha\alpha',n}e^{-i\textbf{k}\cdot\textbf{r}_n},\\
        &\Omega_p(\textbf{k})=\frac{1}{\sqrt{N}}\sum_n\Omega_p(\textbf{r}_n)e^{-i\textbf{k}\cdot\textbf{r}_n},\\
        &\hat{F}_{ge,\textbf{k}}=\frac{1}{\sqrt{N}}\sum_n\hat{F}_{ge}(\textbf{r}_n)e^{-i\textbf{k}\cdot\textbf{r}_n},\\
        &\textbf{G}(\omega,\textbf{k})=\sum_{n}\textbf{G}(\omega,\textbf{r}_{n})e^{-i\textbf{k}\cdot\textbf{r}_n},\\
        &V(\textbf{k})=\frac{1}{N}\sum_{n\neq 0}V(\textbf{r}_{n})e^{-i\textbf{k}\cdot\textbf{r}_n}.
\end{split}
\end{equation}
With this transformation Eqs. (\ref{SEOM1}) become
\begin{equation}
    \begin{split}
&\frac{\partial\tilde{\sigma}_{ge,\textbf{k}}}{\partial t}=-\bigg(\frac{\Gamma_{\textbf{k}}+\gamma_{\mathrm{sc}}}{2}+i(\Delta_{\textbf{k}}-\delta_p)\bigg)\tilde{\sigma}_{ge,\textbf{k}}+i\Omega_p(\textbf{k})+i\Omega_c^*\tilde{\sigma}_{gr,\textbf{k}}+\hat{F}_{ge,\textbf{k}},\\
&\frac{\partial\tilde{\sigma}_{gr,\textbf{k}}}{\partial t}=i\delta\tilde{\sigma}_{gr,\textbf{k}}+i\Omega_c\tilde{\sigma}_{ge,\textbf{k}}-\frac{i}{\hbar}\sum_{\textbf{q}}\sum_{\textbf{k}'}V(\textbf{q})\tilde{\sigma}_{gr,\textbf{k}'}^\dagger\tilde{\sigma}_{gr,\textbf{k}'+\textbf{q}}\tilde{\sigma}_{gr,\textbf{k}-\textbf{q}},\\
    \end{split}
\end{equation}
where
\begin{equation}
    \begin{split}
        &\Gamma_{\textbf{k}}=\frac{2}{\hbar}\frac{\omega_p^2}{c^2\epsilon_0}\textbf{d}^*_{ge}\cdot \mathrm{Im}\bigg[\textbf{G}(\omega_p,\textbf{k})\bigg]\cdot\textbf{d}_{ge},\\
        &\Delta_{\textbf{k}}=-\frac{1}{\hbar}\frac{\omega_p^2}{c^2\epsilon_0}\textbf{d}^*_{ge}\cdot \mathrm{Re}\bigg[\textbf{G}(\omega_p,\textbf{k})\bigg]\cdot\textbf{d}_{ge}.\\
    \end{split}
\end{equation}
Or, denoting $\textbf{d}_{ge}=d\textbf{e}_d$, we can write
\begin{equation}
    \begin{split}
        &\frac{\Gamma_{\textbf{k}}}{2}+i\Delta_{\textbf{k}}=-i\frac{3}{2}\gamma_{\mathrm{atom}}\lambda\textbf{e}_d\cdot \textbf{G}(\omega_p,\textbf{k})\cdot\textbf{e}_d,\quad\mathrm{where}\quad \gamma_{\mathrm{atom}}=\frac{|d|^2\omega_{p}^3}{3\pi\epsilon_0\hbar c^3},\quad\mathrm{and}\quad\lambda=2\pi c/\omega_p.
    \end{split}
\end{equation}
Assuming that the decay rate $\Gamma_{\textbf{k}}$ of $\tilde{\sigma}_{ge,\textbf{k}}$ is faster than the relevant dynamics, $\Gamma_{\textbf{k}}\gg\frac{|\dot{\tilde{\sigma}}_{gr,\textbf{k}}|}{|\tilde{\sigma}_{gr,\textbf{k}}|}$, we can adiabatically eliminate $\tilde{\sigma}_{ge,\textbf{k}}$ from the dynamics. Thus, finding the steady state solution of $\tilde{\sigma}_{ge,\textbf{k}}$
\begin{equation}
    \begin{split}
        &\tilde{\sigma}_{ge,\textbf{k}}=\frac{\Omega_p({\textbf{k}})+\Omega_c^*\tilde{\sigma}_{gr,\textbf{k}}-i\hat{F}_{ge,\textbf{k}}}{\Delta_{\textbf{k}}-\delta_p-i\frac{\Gamma_{\textbf{k}}+\gamma_{\mathrm{sc}}}{2}},
    \end{split}
\label{Ssge}
\end{equation}
and inserting it in the equation of the collective Rydberg modes $\tilde{\sigma}_{gr,\textbf{k}}$ (which from now on will be denoted as $\hat{\sigma}_{\textbf{k}}\equiv \tilde{\sigma}_{gr,\textbf{k}}$), we obtain
\begin{equation}
    \begin{split}
&\frac{\partial\hat{\sigma}_{\textbf{k}}}{\partial t}=-\bigg(\frac{\gamma_{\textbf{k}}}{2}+i(\omega_{\textbf{k}}-\delta)\bigg)\hat{\sigma}_{\textbf{k}}-\Omega_{\textbf{k}}-\hat{F}_{\textbf{k}}-\frac{i}{\hbar}\sum_{\textbf{q}}\sum_{\textbf{k}'}V(\textbf{q})\hat{\sigma}_{\textbf{k}'}^\dagger\hat{\sigma}_{\textbf{k}'+\textbf{q}}\hat{\sigma}_{\textbf{k}-\textbf{q}},\\
    \end{split}
\label{SEOM2}
\end{equation}
with
\begin{equation}
\begin{split}
    &\frac{\gamma_{\textbf{k}}}{2}+i\omega_{\textbf{k}}=\frac{|\Omega_c|^2}{\frac{\Gamma_{\textbf{k}}+\gamma_{\mathrm{sc}}}{2}+i(\Delta_{\textbf{k}}-\delta_p)},\\
    &\Omega_{\textbf{k}}+\hat{F}_{\textbf{k}}=\frac{\Omega_c}{\frac{\Gamma_{\textbf{k}}+\gamma_{\mathrm{sc}}}{2}+i(\Delta_{\textbf{k}}-\delta_p)}\bigg(\Omega_p(\textbf{k})-i\hat{F}_{ge,\textbf{k}}\bigg).\\
\end{split}
\end{equation}
It can be seen that the condition under which our assumption of a fast decaying $\tilde{\sigma}_{ge,\textbf{k}}$ is valid, is $\frac{\gamma_{\textbf{k}}}{2},\omega_{\textbf{k}}-\delta,\Omega_{\textbf{k}}\ll\Gamma_{\textbf{k}}$.
Considering the commutation relations in the linearized regime, $[\hat{\sigma}_{\mathbf{k}},\hat{\sigma}_{\mathbf{k}'}^{\dag}]\approx \delta_{\mathbf{k},\mathbf{k}'}$, where the collective Rydberg modes appear as bosons, the first two terms of Eq. (\ref{SEOM2}) can be seen as resulting from an effective, non-Hermitian Hamiltonian
\begin{equation}
    \begin{split}
        &\hat{H}^{\mathrm{eff}}_{\mathbf{k}}=\hbar \bigg(\omega_{\mathbf{k}}-\delta-i\frac{\gamma_{\mathbf{k}}}{2}\bigg)\hat{\sigma}_{\mathbf{k}}^{\dag}\hat{\sigma}_{\mathbf{k}}
-\hbar\bigg(i\Omega_{\mathbf{k}}\hat{\sigma}_{\mathbf{k}}^{\dag}+\mathrm{h.c.}\bigg),
    \end{split}
    \label{SH_k}
\end{equation}
as in Eq. (3) of the main text.
\subsection{1.3. Input-Output relation of the EM field}
Using the same Heisenberg-Langevin approach, we solve for the electromagnetic (EM) field, obtaining the input-output relation in momentum space (projected onto the transition-dipole orientation $\textbf{e}_{d}$)
\begin{equation}
\begin{split}
     &\hat{E}_{\textbf{k}}(z)={\hat{E}_{0,\textbf{k}}}(z)+E_{p,\textbf{k}}e^{ik_zz}+\frac{\omega_p^2}{c^2\epsilon_0}\textbf{e}_{d}^*\cdot\textbf{G}(\omega_p,z,\textbf{k})\cdot\textbf{e}_{d}d\frac{\tilde{\sigma}_{ge,\textbf{k}}}{\sqrt{N}}.
     \label{Sinout0}
\end{split}
\end{equation}
Here $\hat{E}_{\textbf{k}}(z)$ is the slowly varying envelope of the EM field in $\textbf{k}$ space around carrier frequency $\omega_p$, defined via $\textbf{e}_{d}^*\cdot\hat{\textbf{E}}(\textbf{r})e^{i\omega_pt}=\hat{E}(\textbf{r}_\perp,z)=\sum_{\textbf{k}}\hat{E}_{\textbf{k}}(z)e^{i\textbf{k}\cdot\textbf{r}}$, where $\hat{\textbf{E}}(\textbf{r})$ is the quantum field from Eq. (\ref{SE}), and we add to it the coherent input component $E_p(\textbf{r})$. The corresponding input quantum vacuum field ${\hat{E}_{0,\textbf{k}}}(z)$ is given by a similar definition, using the quantum field from Eq. (\ref{SE}) with the replacement $\hat{a}_{\mathbf{k}k_z\mu}(t)\rightarrow \hat{a}_{\mathbf{k}k_z\mu}(0) e^{-i\nu_{\mathbf{k}k_z}t}$. The coherent input field in $\textbf{k}$ space is $E_{p,\textbf{k}}=\frac{\hbar}{d^*\sqrt{N}}\Omega_p(\textbf{k})$, and $k_z=\sqrt{(\omega_p/c)^2-|\textbf{k}|^2}$. Finally, $\textbf{G}(\omega_p,z,\textbf{k})$ is the (tensor) Green's function at position $z$ away from the $xy$ plane. For a subwavelength lattice, $a<\lambda$, where a single diffraction order exists in the scattering (at least for paraxial illumination), and assuming that the atoms are polarized in the $xy$ plane, one obtains (for $|z|>a$) $\frac{1}{\hbar}\frac{\omega_p^2}{c^2\epsilon_0}\textbf{e}_{d}^*\cdot\textbf{G}(\omega_p,z,\textbf{k})\cdot\textbf{e}_{d}|d|^2=i\frac{\Gamma_{\textbf{k}}}{2}e^{ik_z|z|}$ (see Supplementary Material of Ref.\cite{Scoop}). Inserting Eq. (\ref{Ssge}) into Eq. (\ref{Sinout0}) we obtain
\begin{equation}
\begin{split}
     &\hat{E}_{\textbf{k}}(z)={\hat{E}_{0,\textbf{k}}}(z)+E_{p,\textbf{k}}(0)e^{ik_zz}+r_{\textbf{k}}E_{p,\textbf{k}}(0)e^{ik_z|z|}+r_{\textbf{k}}\hat{E}_{0,\textbf{k}}(0)e^{ik_z|z|}+r_{\textbf{k}}E_c^*e^{ik_z|z|}\hat{\sigma}_{\textbf{k}},
     \label{Sinout1}
\end{split}
\end{equation}
where $r_{\textbf{k}}=-\frac{\frac{\Gamma_{\textbf{k}}}{2}}{\frac{\Gamma_{\textbf{k}}+\gamma_{\mathrm{sc}}}{2}+i(\Delta_{\textbf{k}}-\delta_p)}$, and $E_c=\frac{\hbar\Omega_c}{\sqrt{N}d}$. The total EM field at point $z$ can be decomposed into two parts $\hat{E}_{\textbf{k},\pm}$, a right $(+)$ and left $(-)$ propagating fields, at $z>0$ and $z<0$, respectively. With this notation we obtain
\begin{equation}
\begin{split}
     &\hat{E}_{\textbf{k},\pm}(z)=e^{\pm ik_zz}\bigg((1+r_{\textbf{k}})\hat{E}_{\mathrm{in},\textbf{k},\pm}+r_{\textbf{k}}\hat{E}_{\mathrm{in},\textbf{k},\mp}+r_{\textbf{k}}E_c^*\hat{\sigma}_{\textbf{k}}\bigg),
     \label{Sinout2}
\end{split}
\end{equation}
where $\hat{E}_{\mathrm{in},\textbf{k},\pm}=E_{p,\textbf{k}}\delta_{s,+}+\hat{E}_{0,\textbf{k},\pm}$, for right- ($s=+$) or left- ($s=-$) propagating input field. The Kronecker delta $\delta_{s,+}$ means that the average coherent part of the input light is assumed to be only right-propagating (but vacuum noise is always incident form both sides).
Here, we also used the fact that for a paraxial mode $\textbf{k}$, and within the Markov approximation, we can write $\hat{E}_{0,\textbf{k},\pm}(z)\approx\hat{E}_{0,\textbf{k},\pm}(0)e^{\pm ik_zz}$, although this assumption has no effect on the normal-ordered correlators calculated in Secs. 4 and 5.\\
\subsection{1.4. EIT linear reflectivity}
Without interaction, $V(\textbf{q})=0$, we can solve for the steady state of Eq. (\ref{SEOM2}), and insert it back into Eq. (\ref{Sinout2}), obtaining
\begin{equation}
\begin{split}
     &\hat{E}_{\textbf{k},\pm}(z)=e^{\pm ik_zz}\bigg((1+\tilde{r}_{\textbf{k}})\hat{E}_{\mathrm{in},\textbf{k},\pm}+\tilde{r}_{\textbf{k}}\hat{E}_{\mathrm{in},\textbf{k},\mp}\bigg),
     \label{Sinout3}
\end{split}
\end{equation}
where
\begin{equation}
\tilde{r}_{\textbf{k}}=r_{\textbf{k}}\frac{\delta}{\delta-\omega_{\textbf{k}}+i\gamma_{\textbf{k}}/2}=-\frac{\frac{\Gamma_{\textbf{k}}}{2}\delta}{\bigg(\frac{\Gamma_{\textbf{k}}+\gamma_{\mathrm{sc}}}{2}+i(\Delta_{\textbf{k}}-\delta_p)\bigg)\delta+i|\Omega_c|^2}=-r_{\mathrm{res},\textbf{k}}\frac{1}{1+ir_{\mathrm{res},\textbf{k}}\bigg(2\frac{\Delta_{\textbf{k}}-\delta_p}{\Gamma_{\textbf{k}}}+\frac{\gamma^{(0)}_\textbf{k}}{2\delta}\bigg)},
\end{equation}
with $r_{\mathrm{res},\textbf{k}}=\frac{\Gamma_{\textbf{k}}}{\Gamma_{\textbf{k}}+\gamma_{\mathrm{sc}}}$ being the on-resonance reflectivity magnitude, and $\gamma^{(0)}_\textbf{k}=4|\Omega_c|^2/\Gamma_{\textbf{k}}$. We note that the modified reflectivity $\tilde{r}_{\textbf{k}}$ is maximal at the resonance condition $\frac{\Delta_{\textbf{k}}-\delta_p}{\Gamma_{\textbf{k}}}=\frac{\gamma^{(0)}_\textbf{k}}{4\delta}$, yielding a purely real value $\tilde{r}_{\textbf{k}}=-r_{\mathrm{res},\textbf{k}}$. This is in analogy to the pure absorption in three-level atoms \cite{SEITrev}, and in general does not require single-photon resonance $\delta_p=\Delta_{\textbf{k}}$.
\section{2. Rydberg blockade: reduced Hilbert space}
Here we provide the conditions under which the Hilbert space is blockaded to contain up to single collective excitations, and discuss the relevant dynamics in this reduced space.
\subsection{2.1. Conditions for Rydberg blockaded space}
We first find conditions under which all states in the Hilbert space containing more than single excitations are effectively unpopulated and irrelevant. For simplicity of presentation, we consider the case of weak incident field and that in addition the collective decay can be approximately taken to be independent of $\mathbf{k}$, $\gamma_{\textbf{k}}\approx\gamma$ and $\omega_{\textbf{k}}\approx \omega$. The latter is a good approximation for paraxial light, where the narrow bandwidth in $\mathbf{k}$-space around $\mathbf{k}=0$ is smaller than that of the functions $\gamma_{\textbf{k}}$ and $\omega_{\textbf{k}}$. Then, we can transform Eq. (\ref{SEOM2}) back to real space
\begin{equation}
    \begin{split}
       &\frac{\partial {\hat{\sigma}}_n}{\partial t}=-\bigg(\frac{\gamma}{2}+i(\omega-\delta)\bigg){\hat{\sigma}}_n-\Omega({\textbf{r}_n})-\hat{F}(\textbf{r}_n)-\frac{i}{\hbar}\sum_{m\neq n}V(\textbf{r}_{nm})\hat{\sigma}_m^\dagger\hat{\sigma}_m{\hat{\sigma}}_n,
    \end{split}
\end{equation}
where the corresponding non-Hermitian Hamiltonian becomes
\begin{equation}
    \begin{split}
        &\hat{H}=-\sum_n\hbar\bigg(i\frac{\gamma}{2}-(\omega-\delta)\bigg){\hat{\sigma}}_n^\dagger{\hat{\sigma}}_n
        -i\hbar\sum_n\bigg[\Omega({\textbf{r}_n}){\hat{\sigma}}_n^\dagger-\Omega^*({\textbf{r}_n}){\hat{\sigma}}_n\bigg]
        +\sum_{n>m}V(\textbf{r}_{nm}){\sigma}_n^\dagger{\hat{\sigma}}_n {\hat{\sigma}}_m^\dagger{\hat{\sigma}}_m.
    \label{SHr}
    \end{split}
\end{equation}
For weak fields, $\Omega(\mathbf{r}_n)\ll \gamma$, we can neglect quantum noise/jumps and solve the Schr\"{o}dinger equation of the non-Hermitian Hamiltonian (\ref{SHr}). This means we have to solve for the coefficients of a general state, taken here up to two excitations consistent with the weak field assumption,
\begin{equation}
\begin{split}
    &\ket{\psi}=c_0\ket{0}+\sum_nc_{1,n}\ket{1,n}+\sum_{n>m}c_{2,nm}\ket{2,nm}.\\
\end{split}
\end{equation}
Here $\ket{0}=|g...g\rangle$ is the total ground state, $\ket{1,n}=\hat{\sigma}_n^{\dag}|0\rangle$ is a singly-excited state at atom $n$, and $\ket{2,nm}=\hat{\sigma}_n^{\dag}\hat{\sigma}_m^{\dag}|0\rangle$ is the doubly-excited state at atoms $n$ and $m$.
Solving the Schr\"{o}dinger equation to lowest relevant orders in $\Omega$ we find the steady-state solution
\begin{equation}
    \begin{split}
        &c_0=1,\\
        &c_{1,n}=-\frac{\Omega({\textbf{r}_n})}{\frac{\gamma}{2}+i(\omega-\delta)},\\
        &c_{2,nm}=\frac{\Omega({\textbf{r}_n})\Omega({\textbf{r}_m})}{\bigg(\frac{\gamma}{2}+i(\omega-\delta)\bigg)^2}\frac{1}{1+\frac{i}{\hbar}\frac{V(\textbf{r}_{nm})}{2\bigg(\frac{\gamma}{2}+i(\omega-\delta)\bigg)}}=c_{1,n}c_{1,m}\frac{1}{1+\frac{i}{\hbar}\frac{V(\textbf{r}_{nm})}{2\bigg(\frac{\gamma}{2}+i(\omega-\delta)\bigg)}}.
    \end{split}
\end{equation}
It is evident that as long as $V(\textbf{r}_{nm})\gg|\frac{\gamma}{2}+i(\omega-\delta)|$ the probability amplitude of two excitations goes to zero $c_{2,nm}\rightarrow0$, and multiply-excited states are irrelevant. For this to be true for all the array atoms, we then require that this condition is met for the furthest possible atoms (maximal $\mathbf{r}_{nm}$). This amounts to the requirement that the array linear size is smaller than the relevant blockade radius $R_b$ defined (at resonance $\omega_\textbf{k}=\delta$) as $V(R_b)=\gamma_{\textbf{k}}$. In practice however, the array size is not relevant, but rather the furthest distance between atoms that are actually excited, namely, which are within the beam waist $w$ of the exciting probe light. This yields the condition $w<R_b$.
\subsection{2.2. Dynamics in the reduced Hilbert space}
Back to collective modes $\mathbf{k}$, the blockade mechanism reduces the Hilbert space of our problem to contain only the collective ground state $|0\rangle$ and single excitations of the collective Rydberg modes $|1\rangle_{\mathbf{k}}=\hat{\sigma}_{\mathbf{k}}^{\dag}|0\rangle$. The dynamics are then fully described by
\begin{equation}
    \begin{split}
        \hat{H}^{\mathrm{eff}}=\sum_{\mathbf{k}}\hat{H}^{\mathrm{eff}}_{\mathbf{k}},
    \end{split}
\end{equation}
where $\hat{H}^{\mathrm{eff}}_{\mathbf{k}}$ is the effective Hamiltonian from Eq. (\ref{SH_k}), with the mode operators in the reduced space given by $\hat{\sigma}_{\textbf{k}}=|0\rangle \langle 1|_{\mathbf{k}}$.  We note that $V(\textbf{k})$ does not appear in the Hamiltonian, as the effect of the vdW interaction is fully captured by the fact that our Hilbert space is restricted to include only single excitations. With this effective Hamiltonian we solve the dynamics using the master equation for the density matrix of the atomic collective Rydberg modes,
\begin{equation}
    \begin{split}
        &\frac{\partial \hat{\rho}}{\partial t}=-\frac{i}{\hbar}\bigg(\hat{H}^{\mathrm{eff}}\hat{\rho}-\hat{\rho}(\hat{H}^{\mathrm{eff}})^\dagger\bigg)+\sum_{\textbf{k}}\gamma_{\textbf{k}}\sigma_{\textbf{k}}\hat{\rho}\sigma_{\textbf{k}}^\dagger,
    \end{split}
\label{SME}
\end{equation}
For the results presented below, we solve this master equation for the density matrix and hence for atomic correlators. For two-time correlation functions presented below, we obtain simpler results in the weak field limit, which is defined  within the effective Hamiltonian of the collective Rydberg modes (the effective waveguide QED model), as $|\Omega_{\mathbf{k}}|\ll \gamma_{\mathbf{k}}$. In this limit we can neglect the quantum jumps (since the steady state is reached before they become significant) and solve the Schr\"{o}dinger equation under Hamiltonian $\hat{H}^{\mathrm{eff}}$, for the general state
$|\psi\rangle=c_0|0\rangle+\sum_{\mathbf{k}}c_{\mathbf{k}}|1\rangle_{\mathbf{k}}$. When incident light $\Omega_p(\mathbf{k})$ exits only in certain modes $\mathbf{k}$, one can consider only these modes in the calculations as all other modes remain unpopulated.

\section{3. General relations between field correlators and atomic correlators}
Using the input-output relation, Eq. (\ref{Sinout2}), we can write correlation functions of the EM field operators as a function of atomic correlators. Thus, the equal-time first order correlation function is
\begin{equation}
\begin{split}
    &G_{\textbf{k},s}^{(1)}=\langle\hat{E}^\dagger_{\textbf{k},s}(t)\hat{E}_{\textbf{k},s}(t)\rangle=|b_{\textbf{k},s}|^2|E_{p,\textbf{k}}|^2+|r_{\textbf{k}}|^2|E_c|^2\langle\hat{\sigma}_{\textbf{k}}^\dagger(t)\hat{\sigma}_{\textbf{k}}(t)\rangle+\bigg[b_{\textbf{k},s}^*r_{\textbf{k}}E_c^{\ast} E_{p,\textbf{k}}^*\langle\hat{\sigma}_{\textbf{k}}(t)\rangle+\mathrm{c.c.}\bigg].
\end{split}
\label{SG1}
\end{equation}
where $b_{\textbf{k},+}=1+r_{\textbf{k}}$ and $b_{\textbf{k},-}=r_{\textbf{k}}$.
\\
Similarly, the second order correlation function is
\begin{equation}
    \begin{split}
        &G_{\textbf{k}\textbf{k}',s}^{(2)}(\tau)=\langle\hat{E}^\dagger_{\textbf{k},s}(t)\hat{E}^\dagger_{\textbf{k}',s}(t+\tau)\hat{E}_{\textbf{k}',s}(t+\tau)\hat{E}_{\textbf{k},s}(t)\rangle=\\
        &=|b_{\textbf{k},s}|^2|b_{\textbf{k}',s}|^2|E_{p,\textbf{k}}|^2|E_{p,\textbf{k}'}|^2+\\&+\bigg[E_c\bigg(|b_{\textbf{k}',s}|^2b_{\textbf{k},s}r_{\textbf{k}}^*|E_{p,\textbf{k}'}|^2E_{p,\textbf{k}}\langle\hat{\sigma}_{\textbf{k}}^\dagger(t)\rangle+|b_{\textbf{k},s}|^2b_{\textbf{k}',s}r_{\textbf{k}'}^*|E_{p,\textbf{k}}|^2E_{p,\textbf{k}'}\langle\hat{\sigma}_{\textbf{k}'}^\dagger(t+\tau)\rangle\bigg)+\mathrm{c.c.}\bigg]\\
        &+|E_c|^2\bigg(|b_{\textbf{k}',s}|^2|r_{\textbf{k}}|^2|E_{p,\textbf{k}'}|^2\langle\hat{\sigma}_{\textbf{k}}^\dagger(t)\hat{\sigma}_{\textbf{k}}(t)\rangle+|b_{\textbf{k},s}|^2|r_{\textbf{k}'}|^2|E_{p,\textbf{k}}|^2\langle\hat{\sigma}_{\textbf{k}'}^\dagger(t+\tau)\hat{\sigma}_{\textbf{k}'}(t+\tau)\rangle\\&+\bigg[b_{\textbf{k},s}b_{\textbf{k}',s}^*r_{\textbf{k}}^*r_{\textbf{k}'}E_{p,\textbf{k}}E_{p,\textbf{k}'}^*\langle\hat{\sigma}_{\textbf{k}}^\dagger(t)\hat{\sigma}_{\textbf{k}'}(t+\tau)\rangle+\mathrm{c.c.}\bigg]\bigg)+\bigg[b_{\textbf{k},s}b_{\textbf{k}',s}r_{\textbf{k}}^*r_{\textbf{k}'}^*E_{p,\textbf{k}}E_{p,\textbf{k}'}(E_c)^2\langle\hat{\sigma}_{\textbf{k}}^\dagger(t)\hat{\sigma}_{\textbf{k}'}^\dagger(t+\tau)\rangle+\mathrm{c.c.}\bigg]\\&+\bigg[|E_c|^2E_c^*\bigg(b_{\textbf{k}',s}^*r_{\textbf{k}'}|r_{\textbf{k}}|^2E_{p,\textbf{k}'}^*\langle\hat{\sigma}_{\textbf{k}}^\dagger(t)\hat{\sigma}_{\textbf{k}'}(t+\tau)\hat{\sigma}_{\textbf{k}}(t)\rangle+b_{\textbf{k},s}^*r_{\textbf{k}}|r_{\textbf{k}'}|^2E_{p,\textbf{k}}^*\langle\hat{\sigma}_{\textbf{k}'}^\dagger(t+\tau)\hat{\sigma}_{\textbf{k}'}(t+\tau)\hat{\sigma}_{\textbf{k}}(t)\rangle\bigg)+\mathrm{c.c.}\bigg]\\
        &+|r_{\textbf{k}}|^2|r_{\textbf{k}'}|^2|E_c|^4\langle\hat{\sigma}_{\textbf{k}}^\dagger(t)\hat{\sigma}_{\textbf{k}'}^\dagger(t+\tau)\hat{\sigma}_{\textbf{k}'}(t+\tau)\hat{\sigma}_{\textbf{k}}(t)\rangle.
    \end{split}
\end{equation}
Here we used the fact that the input EM field operator at later time $\hat{E}_{\mathrm{in},\textbf{k},s}(t+\tau)$ with $\tau>0$, commutes with the atomic operator at earlier time $\hat{\sigma}_{\textbf{k}}(t)$ due to causality (easily shown within our Heisenberg-Langevin formulation).\\
\section{4. Results: photon correlation functions}
Considering the excitation of two channels $\textbf{k}$ and $\textbf{k}'$, we solve the master equation taking identical parameters $\Delta_{\textbf{k}}$ and $\Gamma_{\textbf{k}}$ for both channels, which yields $\{\omega_{\textbf{k}},\gamma_{\textbf{k}},r_{\textbf{k}},\tilde{r}_{\textbf{k}}\}=\{\omega_{\textbf{k}'},\gamma_{\textbf{k}'},r_{\textbf{k}'},\tilde{r}_{\textbf{k}'}\}$. We also assume weak field for channel $\textbf{k}$, $\bigg|\frac{\Omega_{\textbf{k}}}{\frac{\gamma_\textbf{k}}{2}+i(\omega_\textbf{k}-\delta)|}\bigg|\ll1$. Calculating the atomic correlators, and using the relation to the EM field correlators from the last section, we find that at steady state the first order correlation function is
\begin{equation}
\begin{split}
    &\frac{G_{\textbf{k},s}^{(1)}}{|E_{p,\textbf{k}}|^2}=|b_{\textbf{k},s}|^2+\frac{|r_{\textbf{k}}-\tilde{r}_{\textbf{k}}|^2-[b_{\textbf{k},s}^*(r_{\textbf{k}}-\tilde{r}_{\textbf{k}})+\mathrm{c.c.}]}{1+S_{\textbf{k}'}},\\
    \label{Sresults1}
\end{split}
\end{equation}
where $S_{\textbf{k}'}=\frac{2|\Omega_{\textbf{k}'}|^2}{(\omega_{\textbf{k}'}-\delta)^2+(\gamma_{\textbf{k}'}/2)^2}=2N \frac{|E_{p,\textbf{k}'}|^2}{|\hbar\Omega_c/d|^2}|1-\tilde{r}_{\textbf{k}'}/r_{\textbf{k}'}|^2$ is the Rydberg blockade saturation parameter.\\
In a similar fashion, we obtain for the second order correlation function
\begin{equation}
    \begin{split}
       &\frac{G_{\textbf{k}\textbf{k}',s}^{(2)}}{|E_{p,\textbf{k}}|^2|E_{p,\textbf{k}'}|^2}=|b_{\textbf{k},s}|^4-\frac{2}{1+S_{\textbf{k}'}}\bigg[\bigg(|b_{\textbf{k},s}|^2b_{\textbf{k},s}^*(r_{\textbf{k}}-\tilde{r}_{\textbf{k}})+\mathrm{c.c.}\bigg)-2|b_{\textbf{k},s}|^2|r_{\textbf{k}}-\tilde{r}_{\textbf{k}}|^2\bigg].
        \label{Sresults2}
    \end{split}
\end{equation}
For the case where the field  at mode $\textbf{k}'$ is also weak, $S_{\textbf{k}'}\ll1$, this equation reduces to Eq. (9) in the main text.\\
For the two-time second order correlation function we solve the Schr\"{o}dinger equation for the atomic state under the influence of the relevant non-Hermitian Hamiltonian and to lowest order in $\Omega_{\mathbf{k}}$, which is valid within the weak field limit, $S_{\mathbf{k}},S_{\mathbf{k}'}\ll1$, where the quantum jumps can be neglected, as explained in Sec. 2.2. Within this weak field calculation, we relax the assumption of identical channel parameters $\Delta_\textbf{k}$ and $\Gamma_\textbf{k}$, obtaining
\begin{equation}
    \begin{split}
        &\frac{G_{\textbf{k}\textbf{k}',s}^{(2)}(\tau)}{|E_{p,\textbf{k}}|^2|E_{p,\textbf{k}'}|^2}=|b_{\textbf{k},s}|^2|b_{\textbf{k}',s}|^2-\bigg[|b_{\textbf{k}',s}|^2b_{\textbf{k},s}^*(r_{\textbf{k}}-\tilde{r}_{\textbf{k}})+|b_{\textbf{k},s}|^2b_{\textbf{k}',s}^*(r_{\textbf{k}'}-\tilde{r}_{\textbf{k}'})+\mathrm{c.c.}\bigg]+\\&+|b_{\textbf{k}',s}|^2|r_{\textbf{k}}-\tilde{r}_{\textbf{k}}|^2+|b_{\textbf{k},s}|^2|r_{\textbf{k}'}-\tilde{r}_{\textbf{k}'}|^2+\bigg[b_{\textbf{k},s}b_{\textbf{k}',s}^*(r_{\textbf{k}}-\tilde{r}_{\textbf{k}})^*(r_{\textbf{k}'}-\tilde{r}_{\textbf{k}'})+\mathrm{c.c.}\bigg]\\
        &+\bigg[\bigg(b_{\textbf{k}',s}^*|r_{\textbf{k}}-\tilde{r}_{\textbf{k}}|^2(r_{\textbf{k}'}-\tilde{r}_{\textbf{k}'})+b_{\textbf{k},s}^*|r_{\textbf{k}'}-\tilde{r}_{\textbf{k}'}|^2(r_{\textbf{k}}-\tilde{r}_{\textbf{k}})\\&-b_{\textbf{k},s}^*b_{\textbf{k}',s}^*(r_{\textbf{k}}-\tilde{r}_{\textbf{k}})(r_{\textbf{k}'}-\tilde{r}_{\textbf{k}'})\bigg)\bigg(e^{-\bigg(\frac{\gamma_{\textbf{k}'}}{2}+i(\omega_{\textbf{k}'}-\delta)\bigg)\tau}-1\bigg)+\mathrm{c.c.}\bigg]\\&+|r_{\textbf{k}}-\tilde{r}_{\textbf{k}}|^2|r_{\textbf{k}'}-\tilde{r}_{\textbf{k}'}|^2\bigg|e^{-\bigg(\frac{\gamma_{\textbf{k}'}}{2}+i(\omega_{\textbf{k}'}-\delta)\bigg)\tau}-1\bigg|^2.
         \label{Sresults3}
    \end{split}
\end{equation}
We note that the correlation time is set by $1/\gamma_{\textbf{k}'}$ which is
the collective width of the channel of the subsequently
detected photon.\\
For the case of an input light only at a single channel $\textbf{k}$, we get the same results, Eqs. (\ref{Sresults1})-(\ref{Sresults3}), with the replacement $\textbf{k}'\rightarrow\textbf{k}$.
\section{5. Quantum correlations and entanglement}
We wish to examine the entanglement between $\hat{E}_{\textbf{k},s}(z)$ and $\hat{E}^\dagger_{\textbf{k}',s}(z)$ for $\textbf{k}\neq\textbf{k}'$ at a detector at position $z$. To this end we will calculate the Duan \emph{et al.} \cite{SDuan} inseparability criterion and the two-mode quantum squeezing as described in the following.\\
For two harmonic-oscillator modes $\hat{a}_{\textbf{k}}$ and $\hat{a}_{\textbf{k}'}$, satisfying $[\hat{a}_{\textbf{k}},\hat{a}_{\textbf{k}'}^\dagger]=\delta_{\textbf{k},\textbf{k}'}$, the Duan \emph{et al.} inseparability parameter $D$ is defined as
\begin{equation}
    D=\frac{1}{2}\bigg[\mathrm{Var}(\hat{X}_{\theta})+\mathrm{Var}(\hat{P}_{\theta})\bigg],
\end{equation}
with the two mode quadratures given by
\begin{equation}
    \begin{split}
        &\hat{X}_{\theta}=\frac{1}{\sqrt{2}}(\hat{a}_{\textbf{k}}e^{i\theta}+\hat{a}_{\textbf{k}}^\dagger e^{-i\theta}+\hat{a}_{\textbf{k}'}e^{i\theta}+\hat{a}_{\textbf{k}'}^\dagger e^{-i\theta})),
    \end{split}
\end{equation}
and
\begin{equation}
    \begin{split}
        &\hat{P}_{\theta}=\frac{i}{\sqrt{2}}(\hat{a}_{\textbf{k}}e^{i\theta}-\hat{a}_{\textbf{k}}^\dagger e^{-i\theta}-\hat{a}_{\textbf{k}'}e^{i\theta}+\hat{a}_{\textbf{k}'}^\dagger e^{-i\theta})).
    \end{split}
\end{equation}
The Duan \emph{et al.} criterion states that if $D<1$ (for a given angle $\theta$), then the two modes are entangled \cite{SDuan}.
To use this criterion for $\hat{E}_{\textbf{k},s}(z)$, we first calculate the commutator
\begin{equation}
   \begin{split}
       &\bigg[\hat{E}_{\textbf{k},s}(z),\hat{E}^\dagger_{\textbf{k}',s'}(z')\bigg]=\bigg[i\sum_{sk_z\geq0,\mu}\sqrt{\frac{\hbar\omega_{\textbf{k},k_z}}{2\epsilon_0V}}\textbf{e}_d\cdot\textbf{e}_{\textbf{k}k_z\mu}\hat{a}_{\textbf{k}k_z\mu}e^{ik_zz},-i\sum_{s'k_z'\geq0,\mu'}\sqrt{\frac{\hbar\omega_{\textbf{k}'k_z'}}{2\epsilon_0V}}\textbf{e}_{\textbf{k}'k_z'\mu'}\cdot\textbf{e}_d\hat{a}_{\textbf{k}'k_z'\mu'}^{\dagger}e^{-ik_z'z'}\bigg]\\&=\sum_{sk_z\geq0}\frac{\hbar\omega_{\textbf{k},k_z}}{2\epsilon_0V}e^{ik_z(z-z')}\delta_{\textbf{k},\textbf{k}'}\delta_{s,s'}\sum_{\mu}|\textbf{e}_d\cdot\textbf{e}_{\textbf{k}k_z\mu}|^2\approx\frac{\hbar\omega_p}{2\epsilon_0AL}\sum_{sk_z\geq0}e^{ik_z(z-z')}\delta_{\textbf{k},\textbf{k}'}\delta_{s,s'}.
   \end{split}
\end{equation}
In the last step we took the paraxial approximation and used $V=AL$, where $A$ and $L$ are the quantization area and length, respectively. Denoting the finite bandwidth of the spatial frequencies $k_z$ by $B$ (which becomes finite following the finite temporal bandwidth of our coarse-grained dynamical picture), and taking the continuous limit $\sum_k\rightarrow \frac{L}{2\pi}\int dk$, we obtain
\begin{equation}
    \begin{split}
        &\bigg[\hat{E}_{\textbf{k},s}(z),\hat{E}^\dagger_{\textbf{k}',s'}(z')\bigg]=E_0^2\delta_{z,z'}\delta_{\textbf{k},\textbf{k}'}\delta_{s,s'}.
    \end{split}
\end{equation}
Here $E_0=\sqrt{\frac{\hbar\omega_p}{2\epsilon_0Adz}}$ is the zero-point field of the quasi-localized mode $\hat{E}_{\textbf{k},s}(z)$ with finite length $dz=\frac{2\pi}{B}$ around $z$. Therefore, for the Duan \emph{et al.} parameter $D$ we will use $\hat{a}_{\textbf{k},s}=\hat{E}_{\textbf{k},s}(z)/E_0$.\\ 
Using the input-output relation Eq. (\ref{Sinout2}) we find that in the weak field regime, $S_{\textbf{k}},S_{\textbf{k}'}\ll1$, and for both $s=\pm$, the parameter $D$ minimized over $\theta$ is
\begin{equation}
    \begin{split}
        &D=1-K\frac{2|E_{p,\textbf{k}}E_{p,\textbf{k}'}|}{E_0^2},\quad\quad K=|(r_{\textbf{k}}-\tilde{r}_{\textbf{k}})(r_{\textbf{k}'}-\tilde{r}_{\textbf{k}'})|.
    \end{split}
\end{equation}
Similarly, the two-mode squeezing parameter is defined by
\begin{equation}
    \begin{split}
        \mathcal{V}=\mathrm{min}_\theta\{\mathrm{Var}[\hat{X}_{\theta}]\}.
    \end{split}
\end{equation}
In the weak field regime we find (for both $s=\pm$)
\begin{equation}
    \begin{split}
    \mathcal{V}=1-\frac{1}{E_0^2}\bigg|(r_{\textbf{k}}-\tilde{r}_{\textbf{k}})^2E_{p,\textbf{k}}^2+(r_{\textbf{k}'}-\tilde{r}_{\textbf{k}'})^2E_{p,\textbf{k}'}^2+2(r_{\textbf{k}}-\tilde{r}_{\textbf{k}})(r_{\textbf{k}'}-\tilde{r}_{\textbf{k}'})E_{p,\textbf{k}}E_{p,\textbf{k}'}\bigg|.
    \label{Ssqueezing1}
    \end{split}
\end{equation}
The cross term is coming from two-mode correlations $\langle\hat{E}_{\textbf{k},s}(z)\hat{E}_{\textbf{k}',s}(z)\rangle$ whereas the individual terms come from individual-mode squeezing correlations $\langle\hat{E}_{\textbf{k},s}(z)\hat{E}_{\textbf{k},s}(z)\rangle$.\\
For the case of equal intensities $|E_{p,\textbf{k}}|=|E_{p,\textbf{k}'}|$ we have
\begin{equation}
    \begin{split}
    \mathcal{V}=1-K_{\mathrm{sq}}\frac{|E_{p,\textbf{k}}|^2}{E_0^2}\quad\quad K_{\mathrm{sq}}=\bigg|(r_{\textbf{k}}-\tilde{r}_{\textbf{k}})^2e^{2i\varphi}+(r_{\textbf{k}'}-\tilde{r}_{\textbf{k}'})^2+2(r_{\textbf{k}}-\tilde{r}_{\textbf{k}})(r_{\textbf{k}'}-\tilde{r}_{\textbf{k}'})e^{i\varphi}\bigg|,
    \label{Ssqueezing2}
    \end{split}
\end{equation}
where $e^{i\varphi}=E_{p,\textbf{k}}/E_{p,\textbf{k}'}$ is the relative phase between the fields. In the case of identical parameters for both channels $r_{\textbf{k}}=r_{\textbf{k}'}$, $\tilde{r}_\textbf{k}=\tilde{r}_{\textbf{k}'}$ and choosing $\varphi=\pi/2$ (where only the cross-term in Eq. (\ref{Ssqueezing1}), originated in inter-channel correlation, contributes) we obtain
\begin{equation}
    \begin{split}
        &D=\mathcal{V}=1-K\frac{2|E_{p,\textbf{k}}|^2}{E_0^2},\quad\quad K=|r_{\textbf{k}}-\tilde{r}_{\textbf{k}}|^2=\frac{r_{\mathrm{res},\textbf{k}}^2}{1+4(\delta-\omega_{\textbf{k}})^2/\gamma_{\textbf{k}}^2}.
    \end{split}
\end{equation}
We note that $K\frac{2|E_{p,\textbf{k}}|^2}{E_0^2}=\bigg(|\frac{\hbar\Omega_c}{\sqrt{N}d}|^2/E_0^2\bigg)|r_{\textbf{k}}|^2S_{\textbf{k}}$, is always smaller than $1$, since it is comprised of a product of small terms: the last term is small in the weak field limit assumed here, $S_\textbf{k}\ll1$, the reflectivity satisfies $|r_\textbf{k}|<1$, whereas the first term satisfies $|\frac{\hbar\Omega_c}{\sqrt{N}d}|^2/E_0^2\sim\frac{\gamma^{(0)}}{Bc}\ll1$ (noting that the bandwidth has to exceed the inverse of the course-grained time resolution $\Delta t$, $\gamma^{(0)}\ll1/\Delta t\leq Bc$).
\section{6. Practical considerations}
As a physical realization we can consider for example a 2D array of Rb87 atoms loaded into a square optical lattice as in \cite{SRUI}, with $N=10^2$, one atom per site, and lattice constant $a=532$ nm. An input light with wavelength $\lambda=780$ nm and waist $w<\sqrt{N}a\sim5.3$ $\mu$m, couples the $\ket{g}=\ket{5S_{1/2},F=2,m_F=2}$ to $\ket{e}=\ket{5P_{3/2},F=3,m_F=3}$ transition. A coupling field with wavelength $480$ nm, couples the $\ket{e}\leftrightarrow\ket{r}=\ket{100S_{1/2},J=1/2,m_J=1/2}$ transition.\\
The natural linewidth of the $\ket{g}\leftrightarrow\ket{e}$ transition of a single atom is $\gamma_{\mathrm{atom}}/2\pi\approx6.06$ MHz \cite{SSteck}. At normal incidence the linewidth of the collective excitation is given by $\Gamma_{\textbf{k}=0}=\frac{3}{4\pi}\frac{\lambda^2}{a^2}\gamma_{\mathrm{atom}}$ \cite{Scoop}, so $\Gamma_{\textbf{k}=0}/2\pi\approx3.1$ MHz.\\
At resonance, and for a coupling field Rabi frequency of $\Omega_c/2\pi=0.75$ MHz, the effective decay rate $\gamma_{\textbf{k}=0}$ of the $\ket{g}\leftrightarrow\ket{r}$ transition, is about $\gamma_{\textbf{k}=0}/2\pi\approx0.73$ MHz. Under these conditions we estimate \cite{SARC} a blockade radius  $R_b\approx15.2$ $\mu$m, so the condition $w<R_b$ is easily fulfilled. We note that the obtained effective decay rate $\gamma_{\textbf{k}=0}$ is much slower than the collective decay rate $\Gamma_{\textbf{k}=0}$ of the $\ket{g}\leftrightarrow\ket{e}$ transition. This is consistent with our assumption of fast decaying collective states $\ket{e}$ used to adiabatically eliminate these states in Eq. (\ref{Ssge}). On the other hand, it is much faster than the neglected natural decay rate of the metastable Rydberg state (which is of the order $\gamma_r\sim1$ kHz), as required.


\begin{thebibliography}{}

\bibitem{CHAr2} D. E. Chang, J. S. Douglas, A. Gonz\'{a}lez-Tudela, C.-L. Hung, and H. J. Kimble, Rev. Mod. Phys. \textbf{90}, 031002 (2018).
\bibitem{KIMc} T. Aoki, B. Dayan, E. Wilcut, W. P. Bowen, A. S. Parkins, T. J. Kippenberg, K. J. Vahala, and H. J. Kimble, Nature \textbf{443}, 671 (2006).
\bibitem{RAU1} E. Vetsch, D. Reitz, G. Sagu\'{e}, R. Schmidt, S. T. Dawkins and A. Rauschenbeutel, Phys. Rev. Lett. \textbf{104}, 203603 (2010).
\bibitem{RAU2} D. Reitz, C. Sayrin, R. Mitsch, P. Schneeweiss and A. Rauschenbeutel, Phys. Rev. Lett. \textbf{110}, 243603 (2013).
\bibitem{KIM2} A. Goban, C.-L. Hung, S.-P. Yu,	J. D. Hood,	J. A. Muniz, J. H. Lee,	M. J. Martin,	A. C. McClung,	K. S. Choi,	D. E. Chang, O. Painter	 and H. J. Kimble, Nat. Commun. \textbf{5}, 3808 (2014).
\bibitem{OROZ} J Lee, J. A. Grover, J. E. Hoffman, L. A. Orozco and S. L. Rolston, J. Phys. B: At. Mol. Opt. Phys. \textbf{48}, 165004 (2015).	
\bibitem{LOD} M. Arcari, I. S\"{o}llner, A. Javadi, S. Lindskov Hansen, S. Mahmoodian, J. Liu, H. Thyrrestrup, E. H. Lee, J. D. Song, S. Stobbe, and P. Lodahl,  Phys. Rev. Lett. \textbf{113}, 093603 (2014).
\bibitem{TOM} J. D. Thompson, T. G. Tiecke, N. P. de Leon, J. Feist, A. V. Akimov,
M. Gullans, A. S. Zibrov, V. Vuleti\'{c} and M. D. Lukin, Science \textbf{340}, 1202 (2013).

\bibitem{CHAr1} D. E. Chang, V. Vuleti\'{c}, and M. D. Lukin, Nat. Photon. \textbf{8}, 685 (2014).
\bibitem{DK} L.-M. Duan and H. J. Kimble, Phys. Rev. Lett. \textbf{92}, 127902 (2004).
\bibitem{BARg} H. Zheng, D. J. Gauthier, and H. U. Baranger, Phys. Rev. Lett. \textbf{111}, 090502 (2013).
\bibitem{CHA} D.~E.~Chang, A.~S.~S{\o}rensen, E.~A.~Demler, and M.~D.~Lukin, Nat. Phys. \textbf{3}, 807 (2007).
\bibitem{KIMd} B. Dayan, A. S. Parkins, T. Aoki, E. P. Ostby, K. J. Vahala, and H. J. Kimble, Science \textbf{319}, 1062 (2008).
\bibitem{DAY1} I. Shomroni, S. Rosenblum, Y. Lovsky, O. Bechler, G. Guendelman, and B. Dayan, Science \textbf{345}, 903 (2014).
\bibitem{LUK4} T. G. Tiecke, J. D. Thompson, N. P. de Leon, L. R. Liu, V. Vuleti\'{c}, and M. D. Lukin, Nature \textbf{508}, 241 (2014).
\bibitem{REM1} B. Hacker, S. Welte, G. Rempe, and S. Ritter, Nature \textbf{536}, 193 (2016).
\bibitem{CIR1} C. Sch\"{o}n, E. Solano, F. Verstraete, J. I. Cirac, and M. M. Wolf, Phys. Rev. Lett. \textbf{95}, 110503 (2005).
\bibitem{ALJ3} A. Gonz\'{a}lez-Tudela, V. Paulisch, D. E. Chang, H. J. Kimble, and J. I. Cirac, Phys. Rev. Lett. \textbf{115}, 163603 (2015).
\bibitem{REM2} B. Hacker, S. Welte, S. Daiss, A. Shaukat, S. Ritter, L. Li, and G. Rempe, Nat. Photon. \textbf{13}, 110 (2019).

\bibitem{POLr} K. Hammerer, A. S. S{\o}rensen, and E. S. Polzik, Rev. Mod. Phys. \textbf{82}, 1041 (2010).
\bibitem{FIM} M. Fleischhauer, A. Imamoglu and J. P. Marangos, Rev. Mod. Phys. \textbf{77}, 633 (2005).
\bibitem{FIRr} O. Firstenberg, C. S. Adams, and S. Hofferberth, J. Phys. B: At. Mol. Opt. Phys. \textbf{49}, 152003 (2016).
\bibitem{GOR1} A. V. Gorshkov, J. Otterbach, M. Fleischhauer, T. Pohl, and M. D. Lukin, Phys. Rev. Lett. \textbf{107}, 133602 (2011).
\bibitem{FRI} I. Friedler, D. Petrosyan, M. Fleischhauer, and G. Kurizki, Phys. Rev. A \textbf{72}, 043803 (2005).
\bibitem{FIR} O. Firstenberg, T. Peyronel, Q.-Y. Liang, A. V. Gorshkov, M. D. Lukin and V. Vuletic, Nature \textbf{502}, 71 (2013).

\bibitem{OL1} I.~Bloch, Nat. Phys. \textbf{1}, 23 (2005).
\bibitem{OL2} I. Bloch, J. Dalibard and S. Nascimb\`{e}ne, Nat. Phys. \textbf{8}, pages 267 (2012).

\bibitem{ADM} R. J. Bettles, S. A. Gardiner and C. S. Adams, Phys. Rev. Lett. \textbf{116}, 103602 (2016).
\bibitem{coop} E. Shahmoon, D. S. Wild, M. D. Lukin, and S. F. Yelin, Phys. Rev. Lett. \textbf{118}, 113601 (2017).
\bibitem{RUI} J. Rui, D. Wei, A. Rubio-Abadal, S. Hollerith, J. Zeiher, D. M. Stamper-Kurn, C. Gross, and
I. Bloch, Nature \textbf{583}, 369 (2020).
\bibitem{RUO2} G. Facchinetti, S. D. Jenkins, and J. Ruostekoski, Phys. Rev. Lett. \textbf{117}, 1243601 (2016).
\bibitem{MNZ} M. T. Manzoni, M. Moreno-Cardoner and A. Asenjo-Garcia, J. V. Porto, A. V. Gorshkov, and D. E. Chang, N. J. Phys. \textbf{20}, 083048 (2018).
\bibitem{ZOL1} A. Grankin, P.-O. Guimond, D. V. Vasilyev, B. Vermersch and P. Zoller, Phys. Rev. A \textbf{98}, 043825 (2018).
\bibitem{ZOL2} P.-O. Guimond, A. Grankin, D. V. Vasilyev, B. Vermersch, and P. Zoller, Phys. Rev. Lett. \textbf{122}, 093601 (2019).
\bibitem{HEN} L. Henriet, J. S. Douglas, D. E. Chang, and A. Albrecht, Phys. Rev. A \textbf{99}, 023802 (2019).
\bibitem{AAMO} E. Shahmoon, M. D. Lukin, and S. F. Yelin, \emph{Advances in Atomic, Molecular, and Optical Physics} \textbf{68}, 1 (Elsevier, 2019).
\bibitem{om1} E. Shahmoon, M. D. Lukin, and S. F. Yelin, Phys. Rev. A \textbf{101}, 063833 (2020).
\bibitem{om2} E. Shahmoon, D. S. Wild, M. D. Lukin, and S. F. Yelin, arXiv:2006.01973 (2020).
\bibitem{RUO3} C. D. Parmee and J. Ruostekoski, Phys. Rev. A \textbf{103}, 033706 (2021).
\bibitem{ANA} A. Asenjo-Garcia, M. Moreno-Cardoner, A. Albrecht, H. J. Kimble, and D. E. Chang, Phys. Rev. X \textbf{7}, 031024 (2017).
\bibitem{ANAm} S. J. Masson and A. Asenjo-Garcia, Phys. Rev. Research \textbf{2}, 043213 (2020).
\bibitem{TAY} T. L. Patti, D. S. Wild, E. Shahmoon, M. D. Lukin, and S. F. Yelin, Phys. Rev. Lett. \textbf{126}, 223602 (2021).
\bibitem{CIRw} D. Castells-Graells, D. Malz, C. C. Rusconi, J. I. Cirac, arXiv:2107.10813 (2021).
\bibitem{JAN1} J. Perczel, J. Borregaard, D. E. Chang, H. Pichler, S. F. Yelin, P. Zoller and M. D. Lukin, Phys. Rev. Lett. \textbf{119}, 023603 (2017).
\bibitem{ROB2} R. J. Bettles, J. Minar, C. S. Adams, I. Lesanovsky and B. Olmos, Phys. Rev. A \textbf{96}, 041603(R) (2017).
\bibitem{RITc} D. Plankensteiner, C. Sommer, H. Ritsch, and C. Genes, Phys. Rev. Lett. \textbf{119}, 093601 (2017).
\bibitem{cQED} E. Shahmoon, D. S. Wild, M. D. Lukin and S. F. Yelin, arXiv:2006.01972 (2020).
\bibitem{LUKr} M. D. Lukin, M. Fleischhauer, R. Cote, L. M. Duan, D. Jaksch, J. I. Cirac, and P. Zoller, Phys. Rev. Lett. \textbf{87}, 037901 (2001).
\bibitem{RIV} R. Bekenstein, I. Pikovski, H. Pichler, E. Shahmoon, S. Yelin, and M. Lukin, Nat. Phys. \textbf{16}, 676 (2020).
\bibitem{CIR} Z.-Y. Wei, D. Malz, A. González-Tudela, and J. I. Cirac, Phys. Rev. Research 3, 023021 (2021).
\bibitem{CHA4} M. Moreno-Cardoner, D. Goncalves, and D. E. Chang, arXiv:2101.01936 (2021).
\bibitem{POH4} L. Zhang, V. Walther, K. M{\o}lmer, and T. Pohl, arXiv:2101.11375 (2021).
\bibitem{BRW1} J. Pellegrino, R. Bourgain, S. Jennewein, Y. R. P. Sortais, A. Browaeys, S. D. Jenkins, and J. Ruostekoski, Phys. Rev. Lett. \textbf{113}, 133602 (2014).
\bibitem{BRW2} S. Jennewein, M. Besbes, N. J. Schilder, S. D. Jenkins, C. Sauvan, J. Ruostekoski, J.-J. Greffet, Y. R. P. Sortais, and A. Browaeys, Phys. Rev. Lett. \textbf{116}, 233601 (2016).

\bibitem{SI} see Supplementary Materials, which includes Refs. \cite{SIr1,SIr2}.
\bibitem{SIr1} Steck, Daniel A. "Rubidium 87 D line data" (2001).
\bibitem{SIr2} N. \u{S}ibali\'{c}, J. D. Pritchard, C. S. Adams, and K. J. Weatherill, Comput. Phys. Commun. \textbf{220}, 319 (2017).

\bibitem{PET} P. Lambropoulos and D. Petrosyan, \emph{Fundamentals of Quantum Optics and Quantum Information} (Springer, 2007).
\bibitem{DUA} L.-M. Duan, G. Giedke, J. I. Cirac, and P. Zoller, Phys. Rev. Lett. \textbf{84}, 2722 (2000).

\bibitem{note1} Corrections due to multiple angles contained in very small beam waists are disucssed in \cite{MNZ,CHA4}.

\bibitem{FAN} J.-T. Shen and S. Fan, Phys. Rev. Lett. \textbf{98}, 153003 (2007).
\bibitem{BAR} H. Zheng, D. J. Gauthier, and H. U. Baranger, Phys. Rev. A \textbf{82}, 063816 (2010).
\end{thebibliography}

\begin{thebibliography}{}
\bibitem{Scoop} E. Shahmoon, D. S. Wild, M. D. Lukin, and S. F. Yelin,
Phys. Rev. Lett. \textbf{118}, 113601 (2017).
\bibitem{SChapterOne} E. Shahmoon, M. D. Lukin, and S. F. Yelin, Advances in
Atomic, Molecular, and Optical Physics \textbf{68}, 1 (Elsevier,
2019).
\bibitem{Smotion} E. Shahmoon, D. S. Wild, M. D. Lukin, and S. F. Yelin,
arXiv:2006.01973 (2020).
\bibitem{SEITrev} M. Fleischhauer, A. Imamoglu and J. P. Marangos, Rev.
Mod. Phys. 77, 633 (2005).
\bibitem{SDuan} L.-M. Duan, G. Giedke, J. I. Cirac, and P. Zoller, Phys.
Rev. Lett. \textbf{84}, 2722 (2000).
\bibitem{SRUI} J. Rui, D. Wei, A. Rubio-Abadal, S. Hollerith, J. Zeiher, D. M. Stamper-Kurn, C. Gross, and
I. Bloch, Nature \textbf{583},
369 (2020).
\bibitem{SSteck} D. A. Steck, Rubidium 87D Line Data, available online at http://steck.us/alkalidata (revision 2.2.2, 9 July 2021).
\bibitem{SARC} N. Šibalić, J. D. Pritchard, C. S. Adams, and K. J.
Weatherill, Comput. Phys. Commun. \textbf{220}, 319 (2017).
\end{thebibliography}
\end{document}